\documentclass{aa}
\bibpunct{(}{)}{;}{a}{}{,} % to follow the A&A style
\usepackage{graphicx}
\usepackage{txfonts}
\usepackage{subfigure}
\usepackage{xcolor}

\newcommand{\pd}[2]{\frac{\partial #1}{\partial #2}}

\newcommand{\td}[2]{\frac{\de #1}{\de #2}}

\newcommand{\de}{\mathrm{d}}
\newcommand{\abs}[1]{\vert #1 \vert}
\newcommand{\cs}{c_\mathrm{s}}
\begin{document}

   \title{Formation of a planetary Laplace resonance through migration in an eccentric disk}

   \subtitle{The case of GJ876}

   \author{Nicolas P. Cimerman
          \inst{1}\fnmsep\thanks{\email{npcphys@gmail.com}}
          \and
          Wilhelm Kley\inst{1}
          \and
          Rolf Kuiper\inst{1}
          }

   \institute{Institut f\"ur Astronomie und Astrophysik, Universit\"at T\"ubingen, Auf der Morgenstelle 10, 72076 T\"ubingen, Germany}

   \date{Received Month DD, YYYY; accepted Month DD, YYYY}

% \abstract{}{}{}{}{} 
% 5 {} token are mandatory
 
  \abstract
  % context heading (optional)
  % {} leave it empty if necessary  
   {Orbital mean motion resonances in planetary systems originate from dissipative processes in disk-planet interactions that lead to orbital migration.
   In multi-planet systems that host giant planets, the perturbation of the protoplanetary disk strongly affects the migration of companion
   planets.}
  % aims heading (mandatory)
   {By studying the well-characterized resonant planetary system around GJ 876 we aim to explore which effects shape disk-driven migration in
   such a multi-planet system to form resonant chains.}
  % methods heading (mandatory)
   {We modelled the orbital migration of three planets embedded in a protoplanetary disk using two-dimensional locally isothermal
   hydrodynamical simulations. In order to explore the effect of several disk characteristics we performed a parameter study by varying the
   disk thickness, $\alpha$ viscosity, mass as well as the initial position of the planets.
   Moreover, we have carefully analysed and compared simulations with various boundary conditions at the disk's inner rim.}
  % results heading (mandatory)
   {We find that due to the high masses of the giant planets in this system, substantial eccentricity can be excited in the disk. This results
   in large variations of the torque acting on the outer lower-mass planet, which we attribute to a shift of Lindblad and corotation resonances
   as it approaches the eccentric gap that the giants create. Depending on disk parameters, the migration of the outer planet can be stopped at the gap edge in a non-resonant state. In other models, the outer planet is able to open a partial gap and to circularize
   the disk again, later entering a 2:1 resonance with the most massive planet in the system to complete the observed 4:2:1 Laplace resonance.}
  % conclusions heading (optional), leave it empty if necessary 
%   {We have shown}
   {Disk-mediated interactions between planets due to spiral waves and excitation of disk eccentricity by massive planets
   cause deviations from smooth inward migration of exterior lower mass planets.
   Self-consistent modelling of the disk-driven migration of multi-planet systems is thus mandatory. Constraints can be placed on the
   properties of the disk during the migration phase, based on the observed resonant state of the system.
   Our results are compatible with a late migration of the outermost planet into the resonant chain, when the giant planet pair
   already is in resonance.}
%   {}


   \maketitle
%
%________________________________________________________________

\section{Introduction}
The interaction of planets with the protoplanetary disk of gas and dust in which they are born, allows them to exchange
angular momentum and energy with it. This mechanism provides planets with the ability to change their orbital elements
and migrate radially through the disk. Ever since this was described in the seminal work by \citet{GoldreichTremaine1980}, plenty of numerical and analytical studies have
focused on determining migration rates in different regimes depending on the physical properties of planet and disk
\citep[for a review, see][]{KleyNelson2012}. Even for the case of single planets, this is still an area of active
research. Since such studies have revealed a wide range of possible migration rates, planets in a multi-planet system can migrate
in a convergent fashion, for example if a planet migrates inward at a higher rate than an interior companion.
The mutual gravitational interaction between the planets allows them to become captured into mean-motion
resonances (MMR) that can have a major impact on the resulting orbital architecture.
While most of the exoplanet systems that
are known today are found to reside outside of exact resonances \citep{WinnFabrycky2015}, several systems are in
a resonant state. Some cases of very compact resonant chains exist, as for example in the famous system around
TRAPPIST-1 \citep{Luger2017}, where seven low-mass planets form such a configuration.
While generally, convergent disk-driven migration has been shown to be an effective channel for the formation of 
orbital resonances, the modelling often uses damping prescriptions that are obtained by studying the
migration in hydrodynamical models of single planets.
However, for planet pairs that are made of planets in different mass regimes, for example an interior giant and an
exterior super-Earth, it has been shown that disk-mediated perturbations can change the migration behaviour 
\citep[e.g.][]{PG2012,Baruteau2013}.

The nearby red M-dwarf star Gliese 876 (GJ 876) hosts four known planets, that constitute one of the most prominent and
dynamically interesting systems that we know of to date. A thorough review of the observational history of this planetary system, which spans two decades, is given in the introduction of \citet{NelsonB2016} and we just summarize some
important characteristics of this system. The growing amount of radial-velocity (RV) observations from several instruments
have subsequently revealed the first confirmed extrasolar mean-motion resonance and, with the discovery of the fourth planet, the only known
instance of an extrasolar three-body Laplace resonance \citep{Rivera2010}, which is comprised of the three outer planets.
Using dynamical orbital fits, the state of the system has been constrained with high precision to be engaged in a deep 4:2:1 MMR, exhibiting
libration of the critical resonant angle $\varphi_\mathrm{L} = \lambda_\mathrm{c} - 3 \lambda_\mathrm{b} + 2\lambda_\mathrm{e}$
around zero with a low amplitude of about $30^\circ$\citep{NelsonB2016,Trifonov2017,Millholland2018}.
However, it is still unclear whether the system is in a low-energy state of pure-double apsidal corotation (ACR), where the resonant angles
always librate or a high-energy state of quasi ACR, where they switch between mostly libration and brief phases of circulation, since both are
compatible with the most recent observations \citep{Millholland2018}. For reference, we show the parameters of the latest orbital fit by
\citet{Millholland2018} in Table \ref{table:mill18}.
\begin{table}
\caption{Masses and orbital parameters of all four planets adopted from the best-fit model of \citet{Millholland2018}. More information and uncertainties are given in their Table 4.}
\label{table:mill18}      % is used to refer this table in the text
\centering                          % used for centering table
\begin{tabular}{c c c c c}        % centered columns (4 columns)
\hline\hline                 % inserts double horizontal lines
Planet & $M[M_\oplus]$ & $a\,[\mathrm{au}]$ & $e$ & $P\,[\mathrm{d}]$\\    % table heading 
\hline                        % inserts single horizontal line
   d & $7.55$ & 0.022 & 0.057 & 1.938\\      % inserting body of the table
   c & $265.6$ & 0.136 & 0.257 & 30.10\\      % inserting body of the table
   b & $845.2$ & 0.219 & 0.033 & 61.11\\
   e & $15.6$ & 0.350 & 0.03 & 123.8\\
\hline                                   %inserts single line
\end{tabular}
\end{table}

What makes this triplet of resonant planets even more interesting from a theoretical perspective is the fact that the two inner planets c and b,
which were first discovered, are both giants with mass ratios of about $q_\mathrm{c} \simeq 2 \cdot 10^{-3}$ and $q_\mathrm{b} \simeq
6.5 \cdot 10^{-3}$ with respect to the star, that are equivalent to several Jupiter masses around a solar mass star.
The outermost planet has a mass similar to Neptune and $q_\mathrm{e} \simeq 1.3 \cdot 10^{-4}$.
Planets of such high masses are expected to significantly shape the disk, for example by opening deep gaps, which warrants detailed modelling of
the migration process.

Different to the regular Laplace resonance of the Galilean satellites where $\varphi_\mathrm{L}$ librates around
$180^\circ$, the resonance in the system around GJ 876 is chaotic on relatively short time-scales. This behaviour has been
recently analysed using analytical models and N-body integrations by \citet{Batygin2015}, who suggested that
a purely dissipative evolution by migration in a laminar disk would not be able to produce the chaotic configuration
and stochastic forcing by turbulence might be needed. In a follow-up study, \citet{Marti2016} identified two different regions they refer to as an inner and outer resonant region, where the latter is chaotic on decadal time-scales.

This particular system has been recently used  by \citet{Puranam2018} to study the chaotic eccentricity excitation of innermost planet, the super-Earth (d), which has a possibly low but finite value, in order to restrict its tidal dissipation.
While this planet is not part of the resonant chain, it still experiences strong chaotic excitations from its companion planets.

The fact that this system and its dynamical state are well-characterized makes it an interesting subject for theoretical studies that explore
evolutionary paths that could lead to its formation.
Just like the observations, such theoretical models have evolved with time and become more complex.
By prescribing damping rates $\tau_{e,a}$ to the eccentricity and semi-major axis
to mimic disk-driven migration in their N-body simulations, \citet{Lee2002} investigated the capture of the two giant planets into a 2:1
resonance. From their results, they constrained the ratio of these damping time-scales such that the obtained equilibrium eccentricities
were compatible with the observations.

First two-dimensional hydrodynamical models were presented by \citet{Snellgrove2001} and \citet{Kley2005},
who confirmed the possibility of capture into the observed resonance of the giant pair using both isothermal and radiative disk models.
They also reported that the most massive planet in the system can excite substantial eccentricity in the disk,
which was further investigated for a non-migrating giant in \citet{KleyDirksen2006}.
However, only the most massive planet b was included in the active domain of their simulation, which
could not provide sufficient damping rates to suppress the eccentricity growth of planet c, which orbited inside a central cavity. Further
investigating this issue, \citet{Crida2008} considered the possibility of an inner disk that could provide eccentricity damping by directly
interacting with planet c. They were able to match the observed resonant state of the giant pair quite accurately. Inferring damping
time-scales from their hydrodynamical models allowed them to reproduce their findings using damped N-body models.
Adopting a locally isothermal equation of state in both two- and three-dimensional models, \citet{Andre2016} also
considered migrating giant planets and investigated the formation and maintainability of several orbital resonances.

After previous works were focused on the interaction of giant planets, \citet{Podlewska2009,PG2012} modelled the migration of a super-Earth that
orbits exterior to a giant planet, which is reminiscent of the planet-pair b-e in GJ 876. They found that the convergent migration of
said planets can be halted or even reversed, depending on model parameters.
They attributed this to the fact that, by means of the spiral waves that the giant planet launches, it is able to transport
angular momentum and energy outward, depositing some fraction of it into the disk via shocks close to the super-Earth,
which can then experience a modified positive torque.
Of particular interest to our study, but at odds with the observation of this particular system is the fact,
that most of their models cannot reproduce the 2:1 MMR of the outer planet pair around GJ 876.
Divergent evolution of planet pairs via this mechanism has also been investigated by \citet{Baruteau2013}, mainly
with the goal of explaining the observed resonant offset in the \textit{Kepler} planets.

While above mentioned studies have considered the formation of MMR in a pair of adjacent planets,
in this work, we aim to construct a self-consistent disk
migration model of all three planets that are involved in the Laplace resonance. This will elucidate the role that previously described processes
played during the formation and put constraints on the nature of the disk.

This paper is structured as follows. In Section \ref{sec:setup} we describe the physical setup of our model and the governing equations, followed by an outline of the numerical methods that were used in Section \ref{sec:numerics}.
In Section \ref{sec:ref} we show the results for a reference model that successfully produces a Laplace resonance.
Models with varied disk parameters and boundary treatment are presented in Section \ref{sec:var}.
We discuss our results in Section \ref{sec:discussion}. We summarize our findings and give an outlook on future work
in Section \ref{sec:sum}.

%__________________________________________________________________

\section{Physical setup}
\label{sec:setup}
In this section we describe the physical model of our simulations and discuss the governing equations of the problem at hand. In our models,
we only include the three planets that comprise the Laplace resonance today. Our neglect of the innermost planet in the system which is a
super-Earth should be justified since we expect it to have negligible influence on the dynamics during the gas-rich phase of the disk.

\subsection{Governing equations}

We work in a non-rotating, cylindrical coordinate system ${\vec{r} = (r, \varphi, z)^\mathrm{T}}$ that is centred on a
central star of mass $M_\ast$ and aligned such that
the $z$-direction points along the rotational axis of the star. Furthermore, we assume the circumstellar disk has zero inclination,
thus its mid-plane disk lies at $z=0$, and that the planets are coplanar with the disk.
To reduce the problem to two dimensions, we consider the vertically integrated hydrodynamical equations.
This allows us to write the continuity equation as 
\begin{equation}
	\pd{\Sigma}{t} + \nabla \cdot \left( \Sigma \vec{v} \right) = 0,
\end{equation}
where $\Sigma = \int \rho \,\de z$ is the surface mass density and $\vec{v} = (v_r, v_\varphi)^\mathrm{T}$ the two-dimensional gas velocity.
We further assume the disk to be in vertical hydrostatic equilibrium so we can write the surface density as $\Sigma = \sqrt{2 \pi} H \rho$, where
$H \equiv c_\mathrm{s} \Omega_{\mathrm{K}}^{-1}$ is the vertical pressure scale-height of the disk, given by the local ratio of the
isothermal sound speed $c_\mathrm{s}$ and the Keplerian orbital frequency $\Omega_{\mathrm{K}}$. The Navier-Stokes equation then reads
\begin{equation}
	\pd{\left( \Sigma \vec{v} \right)}{t} + \nabla \cdot \left( \Sigma \vec{v} \otimes \vec{v} - \mathbb{T} \right)
	 = - \nabla P - \Sigma \nabla \Phi + \Sigma \vec{a}_\mathrm{ind},
\end{equation}
where $\mathbb{T}$ is the viscous stress tensor, which accounts for the effects of viscosity. Adopting the $\alpha$-prescription of
\citet{SS1973}, the kinematic viscosity is given by $\nu = \alpha c_\mathrm{s} H$. The vertically integrated gas pressure is given by
$P = \int p \,\de z$. In this study, we adopt a locally isothermal equation of state, such that $P = \Sigma \cs^2$. We represent by $\Phi$ the total gravitational potential $\Phi = \Phi_\ast + \Phi_\mathrm{p}$,
which is the sum of several contributions. Stellar gravity can be accounted for by a Newtonian potential since it will not be in the active
domain of our simulations
\begin{equation}
	\Phi_\ast = -\frac{G M_\ast}{\abs{\vec{r}}},
\end{equation}
 while the planets' potential uses a smoothing length, to avoid the singularity at the planet position and to account for effects of 
 the vertical extent of the disk:
\begin{equation}
	\Phi_\mathrm{p} = - \sum\limits_{i} \frac{G M_i}{\sqrt{\abs{\vec{r} - \vec{r}_i}^2 + (\varepsilon H)^2}},
\end{equation}
where $M_i$ and $\vec{r}_i$ are the mass and location of planet $i$, respectively, and we set the dimensionless parameter
$\varepsilon = 0.6$, following \citet{MuellerKley2012}. The planets are modelled as point masses. Since the origin of our coordinate system does
not lie in the center of mass, but is centred on the star, it is non-inertial and we have to account for indirect terms in the acceleration,
represented as $\vec{a}_\mathrm{ind}$.

\subsection{Initial conditions}
\begin{table}
\caption{Masses and initial orbital parameters of the planets for our reference model. The inner-most and least massive planet d is not included
in the simulations.}             % title of Table
\label{table:init}      % is used to refer this table in the text
\centering                          % used for centering table
\begin{tabular}{c c c c}        % centered columns (4 columns)
\hline\hline                 % inserts double horizontal lines
Planet & $M[M_\ast]$ & $a_0\,[\mathrm{au}]$ & $P\,[\mathrm{d}]$\\    % table heading 
\hline                        % inserts single horizontal line
   c & $2.042 \cdot 10^{-3}$ & 0.3 & 103.8\\      % inserting body of the table
   b & $6.506 \cdot 10^{-3}$ & 0.525 & 240.4\\
   e & $1.313 \cdot 10^{-4}$ & 2.0 & 1788\\
\hline                                   %inserts single line
\end{tabular}
\end{table}

As a first step, we initialize the disk following a power-law distribution in surface density of the form
$\Sigma = \Sigma_0 \left(r / \mathrm{au} \right)^{-1/2}$, where we vary $\Sigma_0$ for different models. Our locally isothermal models 
are such that the aspect ratio of the disk $h \equiv H/r$ is constant, which implies a power-law for the temperature $T \propto r^{-1} $.
Initial gas velocities are set to a value corresponding to slightly sub-Keplerian circular motion that accounts for the radial pressure gradient:
\begin{equation}
	v_{r,0} = 0, \quad v_{\varphi,0} = v_\mathrm{K} \sqrt{1 - 3h^2/2 },
	\label{eq:init}
\end{equation}
where $v_\mathrm{K} = \Omega_\mathrm{K} r$ is the Keplerian orbital speed. To minimize transient effects that originate from introducing the
planets, we perform an initial phase where we ramp up the planet mass over a time $t_\mathrm{ramp}$, according to
\begin{equation}
	M_i(t) = M_i \cdot \begin{cases}
	[1 - \cos(\pi t/t_\mathrm{ramp})]/2 &\text{ for } t < t_\mathrm{ramp},\\
	1 &\text{ else.}
	\end{cases}
\end{equation}
After this ramping phase, we keep the planet masses constant and do not account for any growth, for example by gas accretion.
We fix the planets on a circular orbit for a time $t_\mathrm{fix} > t_\mathrm{ramp}$ to allow the disk to adapt to the presence of the planets.
During this stage, the giant planets are able to clear substantial gaps in the surface density of the disk. It is only after this relaxation
process, that we release the planets. The final planet masses $M_i$ and initial semi-major axes can be found in Table \ref{table:init}.
We note that the outer planet pair is spaced wider in terms of period ratios than the inner pair. This choice has been adopted to accelerate
the capture of the inner pair into the 2:1 MMR.

%__________________________________________________________________

\section{Numerical methods}
\label{sec:numerics}
In this section we describe the numerical tools that we employ to solve the governing equations described above.

\subsection{The hydrodynamical code}
For all our simulations we employed the publicly available (magneto-)hydrodynamics code \texttt{FARGO3D} \citep{FARGO3D} and made use of
its compatibility with both CPU and GPU architectures during our study.
We note that all results presented in this paper were obtained with the CPU version, however.
The code utilizes a second-order upwind method on a staggered grid to solve
the hydrodynamic equations. We also make use of the special \texttt{FARGO} azimuthal advection algorithm that was introduced originally by
\citet{Masset2000}, allowing for larger time-steps and reducing the numerical diffusivity of the scheme.
To evolve the motion of the planets in time, \texttt{FARGO3D} offers a fifth-order Runge-Kutta method that uses the time-step
that is dictated by the hydrodynamical solver.

\subsection{Simulation domain, resolution and units}
Motivated by previous studies, we choose a simulation domain that allows us to model the inner disk inside our active grid. For our reference
models, the radial domain ranges from $r \in [0.03\,\mathrm{au}, 5.0\,\mathrm{au}]$ and is covered using logarithmically increasing radial cell widths. With this choice of the inner radius of the disk we follow \citet{Crida2008}.
In the azimuthal direction, we cover the full range of the disk, $\varphi \in [0, 2\pi]$ with uniform cell spacing.
In this case, the only sensible choice is a periodic boundary condition (BC) in azimuth. Our reference models have a resolution of
$N_r \times N_\varphi = 576 \times 708$, which results in cells that have an aspect ratio close to unity.
As the unit of time, we adopt the natural choice of an orbital period at $1\,\mathrm{au}$ distance from the star, that is
$T_0 = 2 \pi / \Omega_1 =  1.730\,\mathrm{yr}$. 
In this unit, the time during which the planets are fixed is $t_\mathrm{fix} \simeq 316\, T_0$
and the ramping time is $t_\mathrm{ramp} \simeq 30 \,T_0$.
Since we are interested in the long-term evolution of the system, our simulations
typically run for longer than ten thousand $T_0$.
This corresponds to tens of thousands of orbits of the inner planet pair.
In models where the planets migrate past their present location, we stop the simulation.
We express torques in code units $\Gamma_\mathrm{code} = G M_\ast/R_0$, where $R_0 = 1\,\mathrm{au}$ is the unit radius and
$M_\ast = 0.334 \, M_\odot$.
We conduct a resolution study by repeating model \texttt{a4-h.05-.25s} with a doubled
resolution in both dimensions of $N_r \times N_\varphi = 1152 \times 1416$ and we will discuss it in Appendix \ref{sec:AppRes}.

\subsection{Radial boundary conditions}
After conducting first experiments, we realized that the radial boundary conditions (BCs), especially at the inner boundary, deserve some close
attention. Since we aim to allow for accretion, a closed inner boundary is not feasible. A semi-permeable BC which only allows for outflow,
seems by far more realistic. However, due to the high mass ratio of the planets and their close-in orbits,
the indirect acceleration term in our simulations can act to create an artificial cavity in the inner disk, which has been
reported before by \citet{KleyNelson2008}. This is physically unrealistic and can drastically decrease the time-step due to large accelerations and make the computation
of our models unfeasible over the required long time-scales.
For our standard models, we decided to adopt an inner BC that allows for outflow only in combination with a damping prescription
that acts on the cells close to the boundary. Following \citet{dVB2006}, we solve the following differential equation at the boundary.
\begin{equation}
	\td{X}{t} = - \frac{X - X_\mathrm{d}}{\tau} R^2(r),
\end{equation}
where $X$ is the quantity to be damped to its target value $X_\mathrm{d}$ on a characteristic time-scale $\tau$ and $R(r)$ is a linear ramp
function starting at zero at the beginning of the damping zone and going towards unity when approaching the simulation boundary. In our case $\tau = 0.3 \Omega_\mathrm{K}(r)$ and the inner and outer damping zones span $0.03 \,\mathrm{au} \leq r \leq 0.07 \,\mathrm{au}$ and
${4.6 \,\mathrm{au} \leq r \leq 5 \,\mathrm{au}}$, respectively.

At the inner boundary, we only apply this damping to the velocities. We damp $v_r$ towards a viscous outflow velocity
${({v_r})_\mathrm{d} = v_\mathrm{visc} \,\beta  = - 3 \nu/(2 r) \,\beta }$, similar to \citet{KleyNelson2008}, while $v_\varphi$
is damped to its initial sub-Keplerian value, according to \eqref{eq:init}. In our standard models, we set $\beta = 10/3$ and also conduct one simulation with a ten
times larger outflow velocity ($\beta = 100/3$, labelled \texttt{fo} for fast outflow).

We explore the effect of the inner BC and the damping prescription by repeating our reference simulation but adopting other possibilities,
like an undamped outflow boundary or only damping the azimuthal velocity.
The importance of the inner BC has been stressed especially for simulations of disks around binary systems and examined in
detail by \citet{Thun2017}.
The outer radial boundary is always closed and velocities are damped towards their initial sub-Keplerian values to avoid any reflections of
waves. The damping prescription is not applied to the surface density if not explicitly stated.
Results for several boundary options are presented in the Appendix \ref{sec:bound}.

\subsection{Parameter exploration}
\begin{table*}
\caption{Overview of conducted simulations. We give every simulation a name and list its parameters: the viscosity parameter $\alpha$, the disk surface density at $\Sigma_0$, the disk aspect ratio,  the inner boundary
condition (IBC) and the resulting configuration. Abbreviations used for damping at inner BC: viscous outflow (VO), $v_r$-undamped
outflow (UR) and undamped outflow (ND).
}             
\label{table:sims}
\centering          
\begin{tabular}{l c c c c c l l}
\hline\hline       
                      % To combine 4 columns into a single one 
Name & $\alpha$ & $\Sigma_0$ [g\,cm$^{-2}$] & $h$ & IBC & Comment &Result\\ 
\hline                    
   \texttt{a4-h.05} & $10^{-4}$ & 7500 & 0.05 &  VO & & stalled outside of resonance \\
   \texttt{a4-h.05-.5s} & $10^{-4}$ & 3750 & 0.05 & VO &  & scattering e \& d \\
   \texttt{a4-h.05-.25s} & $10^{-4}$ & 1875 & 0.05 & VO & reference model & Laplace resonance\\
   \texttt{a4-h.05-.125s} & $10^{-4}$ & 937.5 & 0.05 & VO & & Laplace resonance\\
   \texttt{a4-h.07} & $10^{-4}$ & 7500 & 0.07 &  VO & & stalled outside of resonance \\
   \texttt{a4-h.07-.5s} & $10^{-4}$ & 3750 & 0.07  & VO & & stalled outside of resonance \\
   \texttt{a4-h.07-.25s} & $10^{-4}$ & 1875 & 0.07 &  VO & & stalled outside of resonance \\
   \texttt{a3-h.05} & $10^{-3}$ & 7500 & 0.05 & VO & & stalled outside of resonance \\
   \texttt{a3-h.05-.5s} & $10^{-3}$ & 3750 & 0.05  & VO & & stalled outside of resonance \\
   \texttt{a3-h.05-.25s} & $10^{-3}$ & 1875 & 0.05  & VO & & stalled outside of resonance \\
   \texttt{a3-h.07} & $10^{-3}$ & 7500 & 0.07  & VO & & stalled outside of resonance \\
   \texttt{a3-h.07-.5s} & $10^{-3}$ & 3750 & 0.07  & VO & & stalled outside of resonance \\
   \texttt{a3-h.07-.25s} & $10^{-3}$ & 1875 & 0.07 & VO & & stalled outside of resonance \\
   \hline
   \texttt{a4-h.05-.5s-ur} & $10^{-4}$ & 3750 & 0.05 & UR & BC discarded as unphysical & stalled outside of resonance \\
   \texttt{a4-h.05-.5s-fo} & $10^{-4}$ & 3750 & 0.05 & VO & $\beta = 100/3$& stalled outside of resonance \\
   \texttt{a4-h.05-.25s-ur} & $10^{-4}$ & 1875 & 0.05  & UR & BC discarded as unphysical & 6:3:1 MMR \\
   \texttt{a4-h.05-.25s-fo} & $10^{-4}$ & 1875 & 0.05   & VO & $\beta = 100/3$ & Laplace resonance\\
      \hline
   \texttt{ae1.5-a4-h.05-.25s} & $10^{-4}$ & 1875 & 0.05   & VO & $a_\mathrm{e,0} = 1.5 \, \mathrm{au}$& Scattering e \& d\\
   \texttt{ae3.0-a4-h.05-.25s} & $10^{-4}$ & 1875 & 0.05   & VO & $a_\mathrm{e,0} = 3.0 \, \mathrm{au}$& Laplace resonance\\
   \texttt{gc-a4-h.05-.25s} & $10^{-4}$ & 1875 & 0.05   & VO & $a_\mathrm{b,0} = 0.35 \, \mathrm{au}$,
   $a_\mathrm{b,0} = 0.2 \, \mathrm{au}$& stalled outside of resonance \\
   \texttt{NG-a4-h.05-.25s} & $10^{-4}$ & 1875 & 0.05   & VO & no PP-interaction of e & stalled outside of resonance \\
   \texttt{HM-a4-h.05-.25s} & $10^{-4}$ & 1875 & 0.05   & VO & higher mass $M_\mathrm{e} = 3 \cdot 10^{-4} M_\ast$ & Laplace resonance \\
   \hline
   \texttt{a2-h.05-.5s} & $10^{-2}$ & 3750 & 0.05   & VO & high viscosity test & divergent migration\\
   \hline
   \texttt{a4-h.05-.25s-ir} & $10^{-4}$ & 1875 & 0.05   & VOC & standard damping + closed BC & stalled close to 2:1 MMR \\
	\texttt{a4-h.05-.25s-nd} & $10^{-4}$ & 1875 & 0.05   & ND & no damping at all & 	\\
\hline                   
\end{tabular}
\end{table*}

In our locally isothermal framework, we explore the influence of several disk parameters on the structure and evolution of the
system. For our standard parameter scan, we vary the magnitude of viscosity, given by $\alpha \in [10^{-4},10^{-3}]$,
which is in the range of values suggested by recent simulations of the turbulence created by the vertical shear instability
\citep[VSI,][]{Stoll2017}.
We explore different values for the disk temperature by varying $h \in [0.05, 0.07]$
and consider three different disk masses by adjusting $\Sigma_0$.

For reference, we note that $\Sigma_0 = 1875\,\mathrm{g\,cm^{-2}}$ corresponds to a disk mass of $M_\mathrm{d} = 2.95 \cdot 10^{-2} \, M_\ast$ inside the domain.
Outside of our standard parameter scan, we performed additional simulations, featuring for example $\alpha = 10^{-2}$, an even lower disk
mass or a higher mass outer planet.
In Table \ref{table:sims}, we list the simulations that we conducted and their parameters.

%__________________________________________________________________

\section{Reference hydrodynamical model}
\label{sec:ref}
In this section we present the results of our reference model labelled \texttt{a4-h.05-.25s}, which resulted in the successful formation of
a Laplace resonance. This choice of reference was only made after knowing of this outcome. We will later refer to this simulation to discuss differences in
the behaviour of models that use other values for physical parameters or numerical boundary conditions.

\subsection{Relaxed disk structure}
\begin{figure*}
   \resizebox{\hsize}{!}
            {
            \includegraphics[]{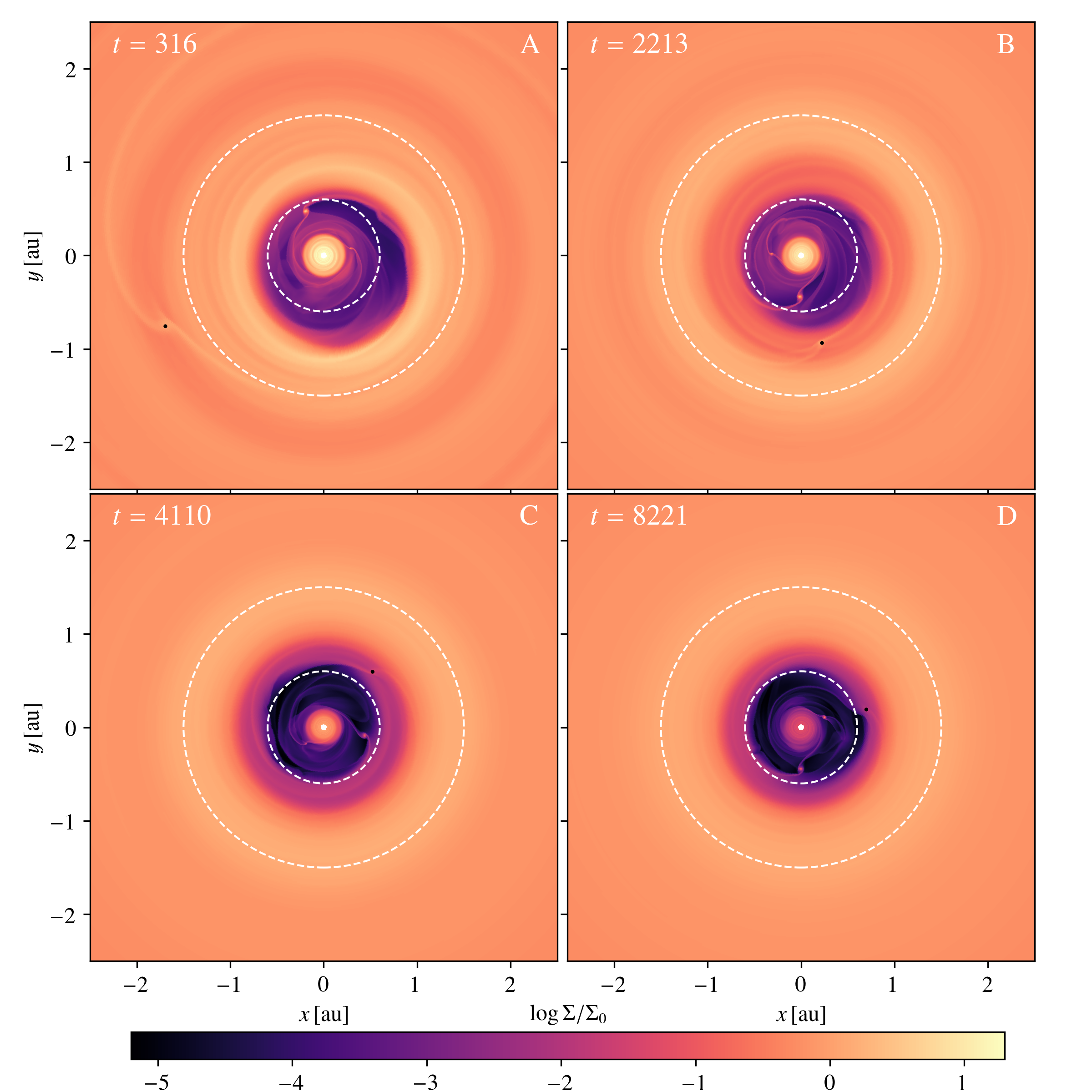}
            }
   \caption{Time series of the surface density for our reference simulation. Time advanced from left to right and top to 	
   bottom. The dashed white circles are of radius $r=0.6 \, \mathrm{au}$ and $r=1.5 \, \mathrm{au}$, respectively and are meant to highlight the
   eccentric nature of the gap. We note that this plot does not show the full simulation domain, which extends further out to $5\,
   \mathrm{au}$. Panel A shows the state at the end of the relaxation process for our reference model.
   A zoomed close-up version of this Figure can be found in the Appendix as Fig. \ref{fig:sig2dRefZoom}.}
   \label{fig:sig2dRef}
\end{figure*}

First, we present the structure of the disk after performing the process of introducing the planets and letting the disk adapt.
The corresponding two-dimensional surface density distribution is presented in panel A (top left) of Fig. \ref{fig:sig2dRef}.
Common to all models is the fact that the giant planets clear a substantial gap, which they share due to their proximity and
high masses. At the outer edge of the gap, located at roughly 1.2 au, a surface density maximum is created.
We note that the gap in which the giant planets are located has gained some eccentricity, as can be seen by comparing its profile to the white
circle. Up to this stage, the planets were still held fixed on a circular orbit.
Notice that the outermost planet was able to carve a partial gap into the disk on its initial circular orbit.
The remaining panels of Fig. \ref{fig:sig2dRef} show how the outer planet migrates inward and settles at the outer gap edge.
There is also a clear evolution of the inner disk and gap which are both depleted in mass as time passes.
During this evolution the eccentricity of the disk is clearly reduced.

\subsection{Planet migration into a Laplace resonance}
\begin{figure*}
   \resizebox{\hsize}{!}
            {               
            \includegraphics[]{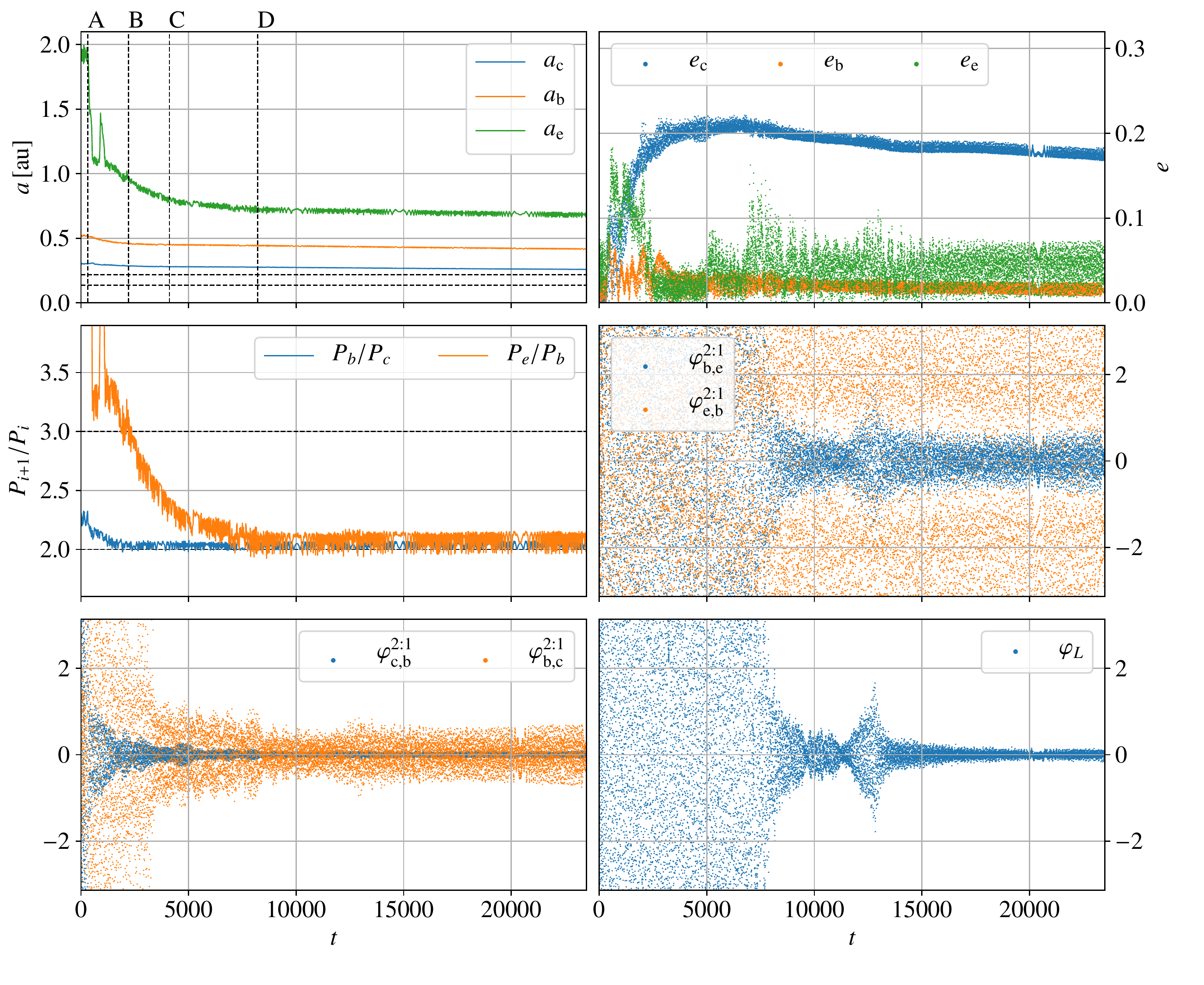}  
            }
   \vspace*{-3em}
   \caption{Time evolution of our reference simulation. From left to right and top to bottom the panels show the semi-major axis $a_i$,
    eccentricity $e_i$, period ratio of adjacent planets $P_{i+1}/P_i$, the resonant angles corresponding to
    the 2:1 resonance of the outer and inner planet pairs $\varphi_{i,j}^{2:1}$ (see e.g. Eqs. \eqref{eq:211}, \eqref{eq:212})
    and the Laplace angle $\varphi_\mathrm{L} = \lambda_\mathrm{c} - 3 \lambda_\mathrm{b} + 2\lambda_\mathrm{e}$.
    In the top left panel, dashed vertical lines mark the times at which Fig. \ref{fig:sig2dRef}
    shows the surface density. The two dashed horizontal lines indicate the semi-major axis of the giant planets as observed.
            }
   \label{fig:MigRef}
\end{figure*}

\label{subsection:stop}
\begin{figure}
   \centering
   \includegraphics[width=.49\textwidth]{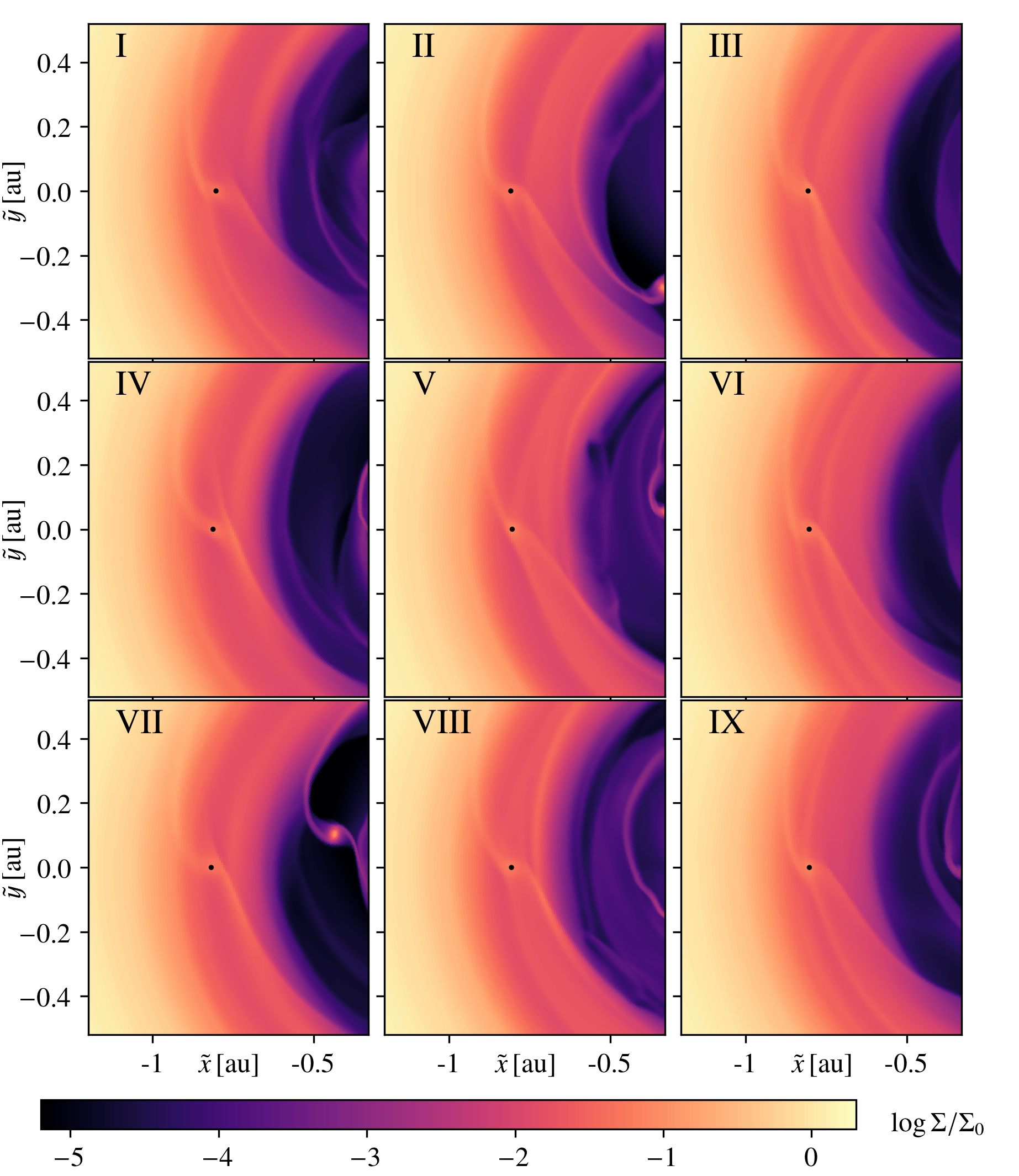}
\\
      \includegraphics[width=.49\textwidth]{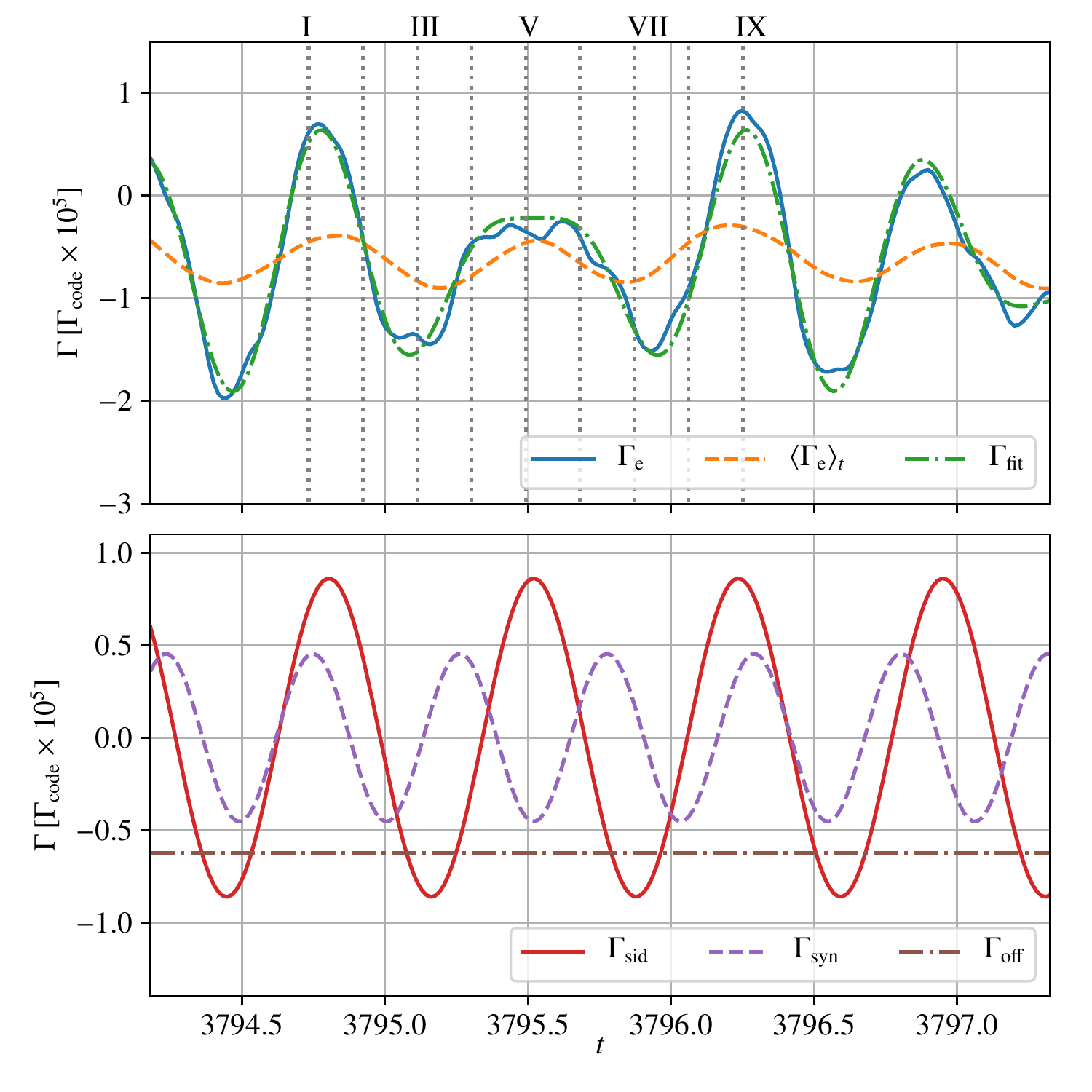}
        \vspace*{-2em}
      \caption{\textit{Top:} Close-up view of the surface density around the Neptune, which is marked by the black dot, for the reference model.
      We use coordinates that are rotated such that planet e is located around $\tilde{y} = 0$
      with $\tilde{x} < 0$. This shows the state where the outer planet is migrating inward
      and located close to the gap edge around $t \simeq 4000$. The Neptune's outer spiral arm remains intact and shows little change,
      the inner spiral arm interacts with the eccentric gap edge. The spiral arm driven by the giant planet can be
      seen passing through the planet's location.
      \textit{Middle:} The blue line shows the torque acting on planet e and the dashed orange line
      shows its moving time average. The green dash-dotted line represents a fit according to Eq. \ref{eq:fittrq}.
      Vertical dotted lines refer by labels to the moments in time that we
      plot in the top panel.
	  \textit{Bottom:} The three components of our fit: the red solid line represents torque variations on the sidereal period of the Neptune,
	  the purple dashed line corresponds to the synodic variations and the brown dash-dotted line represents a constant offset.
              }
                 \label{fig:migin}
\end{figure}

We plot the time evolution of the orbital elements of all three planets in Fig. \ref{fig:MigRef}.
As the planets are freed and allowed to migrate, the giant planets quickly settle into a 2:1 MMR that displays a
low-amplitude libration of the resonant angle
\begin{equation}
	{\varphi^{2:1}_{\mathrm{c},\mathrm{b}} \equiv 2 \lambda_\mathrm{b} - \lambda_\mathrm{c} - \varpi_\mathrm{c}}.
	\label{eq:211}
\end{equation}
During this resonant capture, the eccentricity of the inner-most planet c is observed to increase and continues to do so as the pair migrates jointly, until it settles around a value of $e_\mathrm{c} \simeq 0.2$. Around $t \simeq 3400$, the second
resonant angle
\begin{equation}
{\varphi^{2:1}_{\mathrm{b},\mathrm{c}} \equiv 2 \lambda_\mathrm{b} - \lambda_\mathrm{c} - \varpi_\mathrm{b}}
	\label{eq:212}
\end{equation} switches from
circulation to libration as well. This means that the giant planet pair settles into a state of apsidal corotation (ACR),
where the apsidal lines are aligned and corotate at the same average rate $\dot{\varpi}$, with only small deviations.
The delay in the capture of the critical angles into resonance is related to the initially low eccentricity of the inner planet and a
$1/e$ dependence of the resonance-induced precession rate, as was argued before by \citet{Kley2005}.

The outer Neptune-mass planet (which we will refer to as \textit{the Neptune} from here on)
initially migrates inward in a rapid fashion, such that it is able to approach a 3:1 period ratio with respect to
planet b at around $t \simeq 1500$. This is followed by an increase in the planets eccentricity to about $e = 0.15$ and a phase of rapid outward migration
out to about 1.4 au.
There, migration is reversed again, and the planet is able to become temporarily engaged in the 3:1 orbital resonance with planet
b.
This configuration is maintained only for a very short time until around $t \simeq 2000$, the Neptune is able to escape from 
resonance. During this process, its eccentricity is visibly reduced to values smaller than 0.05.
From this point onward the Neptune migrates inward slowly at the edge of the gap.

As the Neptune reaches the vicinity of a 2:1 MMR with planet e, resonant effects become noticeable in its eccentricity.
At around $t \simeq 8000$, the resonant angle $\varphi^{2:1}_{e,b}$ starts to librate around zero with a low amplitude.
%\ncem{does $\varphi^{2:1}_{e,b}$ show libration around $\pi$?}
Together with the deep resonance of the inner planets, this results in a libration of the
Laplace angle $\varphi_\mathrm{L}$ around zero and completes the 4:2:1 Laplace MMR, similar to the observed configuration.
We have confirmed that only one of the resonant angles of the outer 2:1 MMR librates, which means that this state is not
that of double apsidal corotation.
However, we note that in this model, the migration of the planets was not yet able to drive the planets to their observed semi-major axes.
At the end of the simulation, the innermost planet c is located at $a_\mathrm{c} \simeq 0.27\,\mathrm{au}$,
which is roughly two times the observed value.
Given the time-scales of this rapid assembly of the resonance, the typical lifetime of a protoplanetary disk is of order $\sim$\,Myr, such that
there is plenty of time for the system to become more compact in a slow fashion.
We later discuss a model, where we started the giant planets closer to their observed locations.

We show a time-series of the surface density and the disk torque acting on
planet e during the phase of convergent inward migration at around $t \simeq 3800$ in Fig. \ref{fig:migin}.
In the top panel, the surface density is displayed in a coordinate frame that is rotated,
such that the Neptune is always located around $\tilde{y} \simeq 0$ with $\tilde{x} < 0$.
Clearly, the inner spiral arm and the corotation region interact with the
spiral waves driven by the giant planets and the gap edge, as the planet passes through them.

The torque acting on the Neptune is presented in the middle
panel, with blue being the total torque and the orange dashed line showing a moving average, which remains negative in this
time-window, which is consistent with the observed inward migration. The green dash-dotted line shows a fit which we describe next.

From the surface density plots, we expect that variations of the torque are caused by two main effects: the outer planet experiences variations
to its surrounding surface density on its sidereal period, as it passes through the eccentric gap edge, which we will call $\Gamma_\mathrm{sid}$
. Since the spiral arm driven by the giant planet is corotating with it, the Neptune passes through it on the synodic period of the outer planet
pair $P_\mathrm{syn} = \left(P_\mathrm{b}^{-1} - P_\mathrm{e}^{-1} \right)^{-1}$.
We will refer to this contribution as $\Gamma_\mathrm{syn}$.
To confirm that these two effects make up the observed torque variation, we perform a fit that allows for two sinusoidal components
of different amplitude, period and phase together with a constant offset, according to:
\begin{equation}
	\label{eq:fittrq}
	\Gamma_\mathrm{fit} (t) = \Gamma_\mathrm{off} + \hat{\Gamma}_\mathrm{sid} \sin \left( 2 \pi \frac{t}{P_\mathrm{sid}} + \phi_0 \right) +
	\hat{\Gamma}_\mathrm{syn} \sin \left(2 \pi  \frac{t}{P_\mathrm{syn}} + \phi_1 \right),
\end{equation}
where the second and third term on the right hand side represent $\Gamma_\mathrm{sid}$ and $\Gamma_\mathrm{syn}$, respectively.
The constant $\Gamma_\mathrm{off}$ allows for a net torque, which can be of either sign and is needed since in this very simple approach both
time-depending terms are strictly sinusoidal and have no net contribution.
We note here that this simple fit works as an approximation for short time scales. However, since
the planets migrate, the periods and amplitudes are subject to change.

In the bottom panel of Fig. \ref{fig:migin} we show the three components of Eq. \ref{eq:fittrq}. The periods that we retrieve match the expected
values of  $P_\mathrm{sid}$ and $P_\mathrm{syn}$. Regarding the amplitude of the variations, we find that the eccentricity-related
part dominates the giants wake slightly with an amplitude $\hat{\Gamma}_\mathrm{sid} \simeq 8.61 \cdot 10^{-6} > \hat{\Gamma}_\mathrm{syn}
\simeq 4.54 \cdot 10^{-6}$, but both effects contribute. To match the observed negative net-torque, a negative offset of comparable magnitude
is required $\Gamma_\mathrm{off} \simeq - 6.25 \cdot 10^{-6}$.

This negative offset seems to be caused by the outer, spiral arm that is trailing the Neptune, leading to a negative torque.
From the density plots we see that it is changing only slightly and passes through a region of higher density, which would lead to
a stronger negative contribution of the differential Lindblad torque.
At the location of torque maxima, we recognize regions of higher density leading the planet and might be related to
resonances that are shifted due to the eccentricity of disk and planet, which we will discuss later, as well as interaction with
the giant planet's wake.

\subsection{Evolution of the disk eccentricity}
\begin{figure}
   \centering
   \includegraphics[width=.5\textwidth]{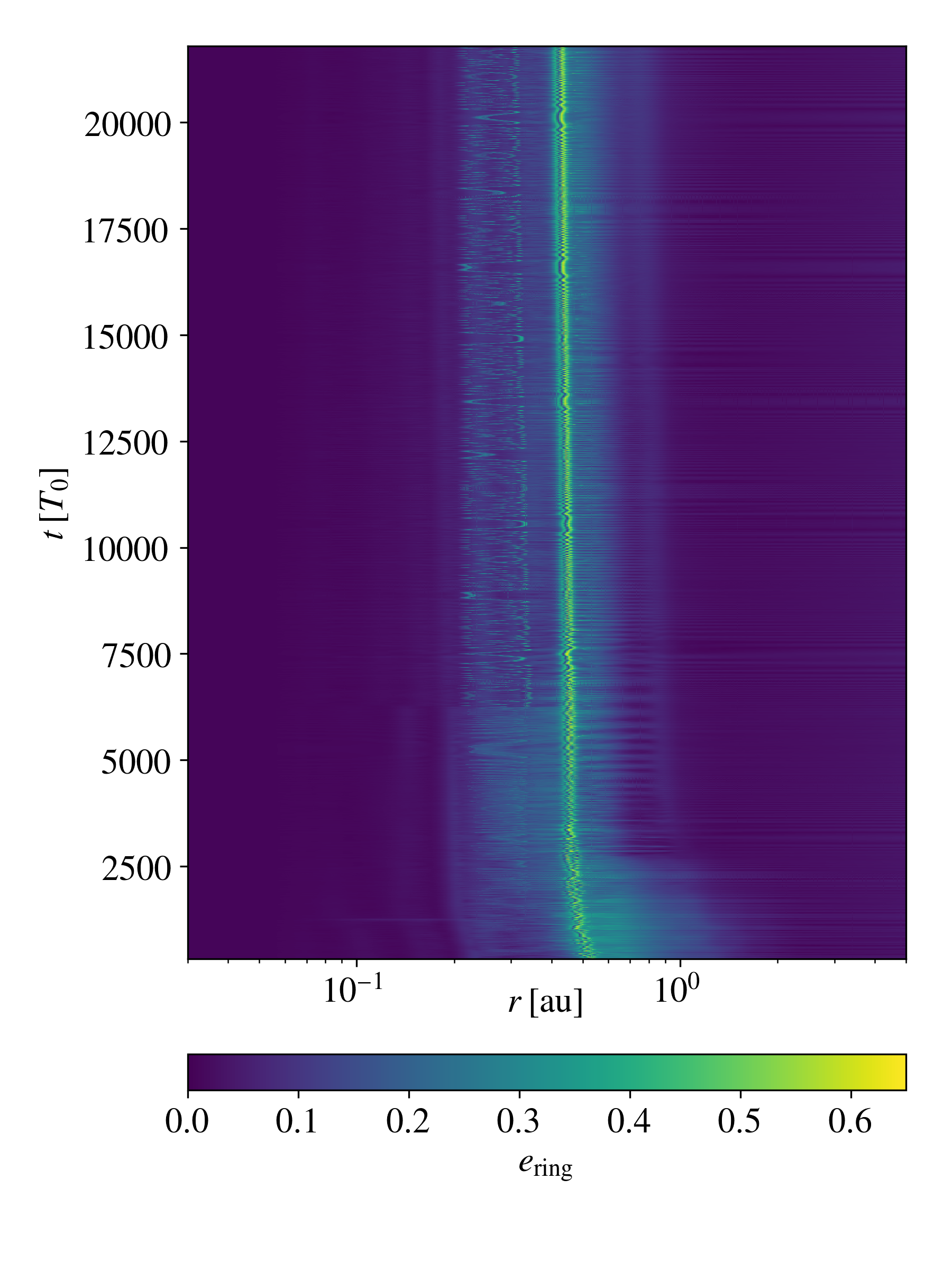}
   	\vspace*{-4em}
      \caption{Time evolution of the eccentricity of rings as defined in \eqref{eq:ering} for the reference simulation,
	  inspired by the plots in \citet{Ragusa2018}. While initially the gap edge $0.6 \lesssim r \lesssim 2.0$ shows substantial eccentricity,
	  this region of the disk is circularized as the planets migrates inward.}
         \label{fig:edisk}
\end{figure}

We now turn to discussing the evolution of the disk eccentricity and the alignment of the disk during the migration of the planets.
For this purpose, we calculate the 
mass-weighted, azimuthally averaged eccentricity of rings, defined in \citet{Thun2017}:
\begin{equation}
	\label{eq:ering}
	e_\mathrm{ring}(r) = \left[ \int\limits_r^{r+\Delta r} \int\limits_0^{2 \pi} \Sigma \, e_\mathrm{cell} r' \de \varphi \,\de r' \right]
	\left[ \int\limits_r^{r+\Delta r} \int\limits_0^{2 \pi} \Sigma \, r' \de \varphi \,\de r' \right]^{-1},
\end{equation}
where $\Delta r$ is the radial cell width of the ring and $e_\mathrm{cell} = \abs{\vec{e}_\mathrm{cell}}$ represents the eccentricity of a test
particle in orbit around the star with the eccentricity vector
\begin{equation}
	\vec{e}_\mathrm{cell} = \frac{\vec{v} \times (\vec{r} \times \vec{v})}{G M_\ast} - \hat{\vec{e}}_r.
\end{equation}
In Fig. \ref{fig:edisk}, we show the time-evolution of the radial profile of this quantity. We note that the time axis begins at the point where
the planets are released. While the inner- and outermost parts of the disk do not show substantial eccentricity, we recognize the eccentric
nature of the gap and disk, that we saw in Fig. \ref{fig:sig2dRef}, as the bright area that extends from roughly
0.5\,au - 1.5\,au in the pre-migration disk state.
As the Neptune approaches the gap edge, the extent of this eccentric feature decreases and as the planet crosses the
3:1 MMR at around $t \simeq 2500$, the disk outside its location relaxes to a more circular state. Another switch in the inner
disk can be recognized as the three planets start becoming engaged in the Laplace resonance at around $t \simeq 6000$.

\subsection{Evolution of the inner disk and gap}
\begin{figure*}
   \resizebox{\hsize}{!}
            {   
   \includegraphics[width=.5\textwidth]{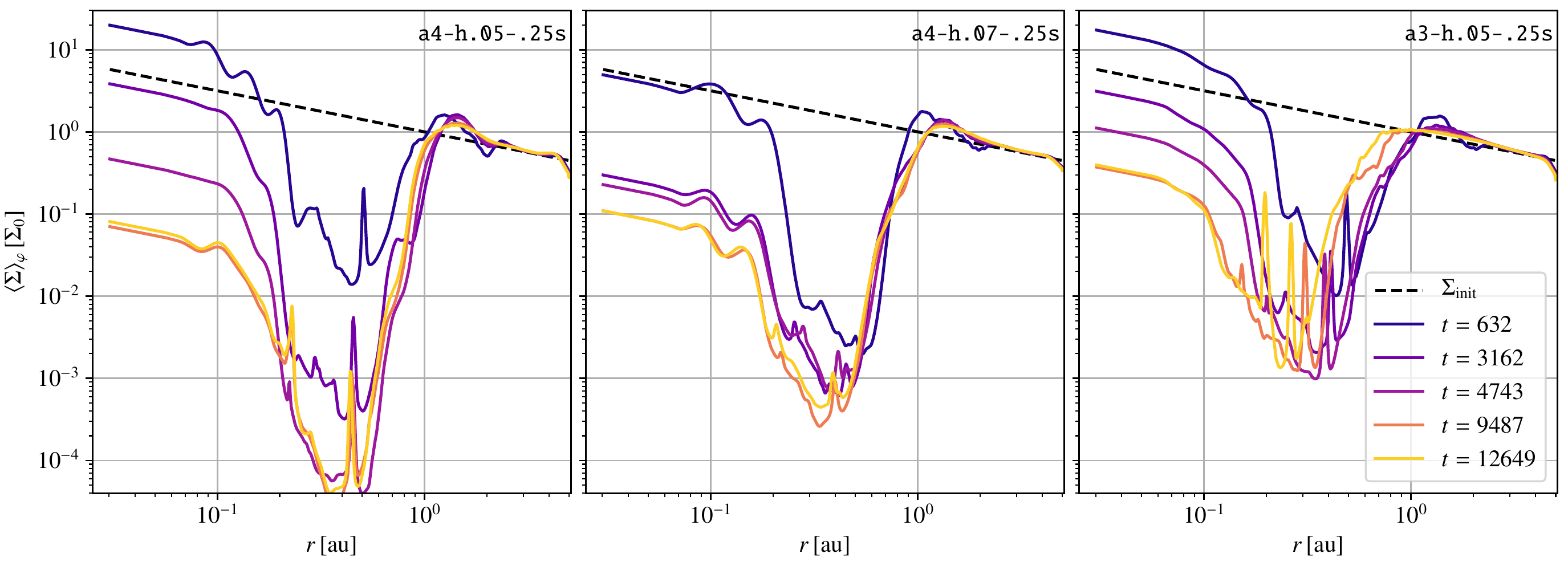}
   	}
   	   	\vspace*{-2em}
      \caption{Time-series of the azimuthally averaged surface density profile for three models. The left panel shows our results for the 
      reference model. Clearly visible is a decrease of the surface density in the inner disk and the gap. The migration of the Neptune into the
		gap can be recognized by the plateau that it creates in the surface density profile at later times.
		The middle and right panels show the same quantity for a model with increased aspect ratio and viscosity, respectively.
              }
         \label{fig:sig2dT}
\end{figure*}

Because of our choice of an outflow boundary condition in combination with enforcing a negative radial velocity at the inner
boundary, that is of order of the viscous velocity, the inner disk is able to viscously drain towards the star.
Torques from the planets and viscous stresses provide means to drive accretion of the inner disk towards the boundary.
The resulting evolution of the inner disk is already visible in Fig. \ref{fig:sig2dRef}. In the left panel of Fig. \ref{fig:sig2dT}, we show a
time-series of the corresponding azimuthally averaged surface density profiles for the reference model, that shows both the radial extent as
well as the surface density of the inner disk decreasing gradually.
From the point where the outer planet migrates past the surface density maximum at the gap edge, this draining
of the gap and inner disk accelerates. This can be understood as a consequence of the Neptune repelling gas from the gap edge and stopping
it from accreting through the gap onto the inner disk. By this effective starving of the gap, wave-mediated outward angular momentum
transport from the giants might become less effective. The observed decrease of the disk eccentricity and the amplitude of torque oscillations
associated with it, are possibly a result of the decreased surface density in the gap. We expect that contributions of the giants' eccentricity
driving resonances are reduced due to a lack of gas and possibly some eccentricity damping is provided by the partially gap-opening Neptune.

%__________________________________________________________________

\section{Variation of disk parameters and boundary condition}
\label{sec:var}
In this section we explore the effect that the variation of the disk thickness, viscosity and mass, as well as the treatment of the
inner boundary condition have on the evolution of the system.

\begin{figure}
   \centering
   \includegraphics[width=.5\textwidth]{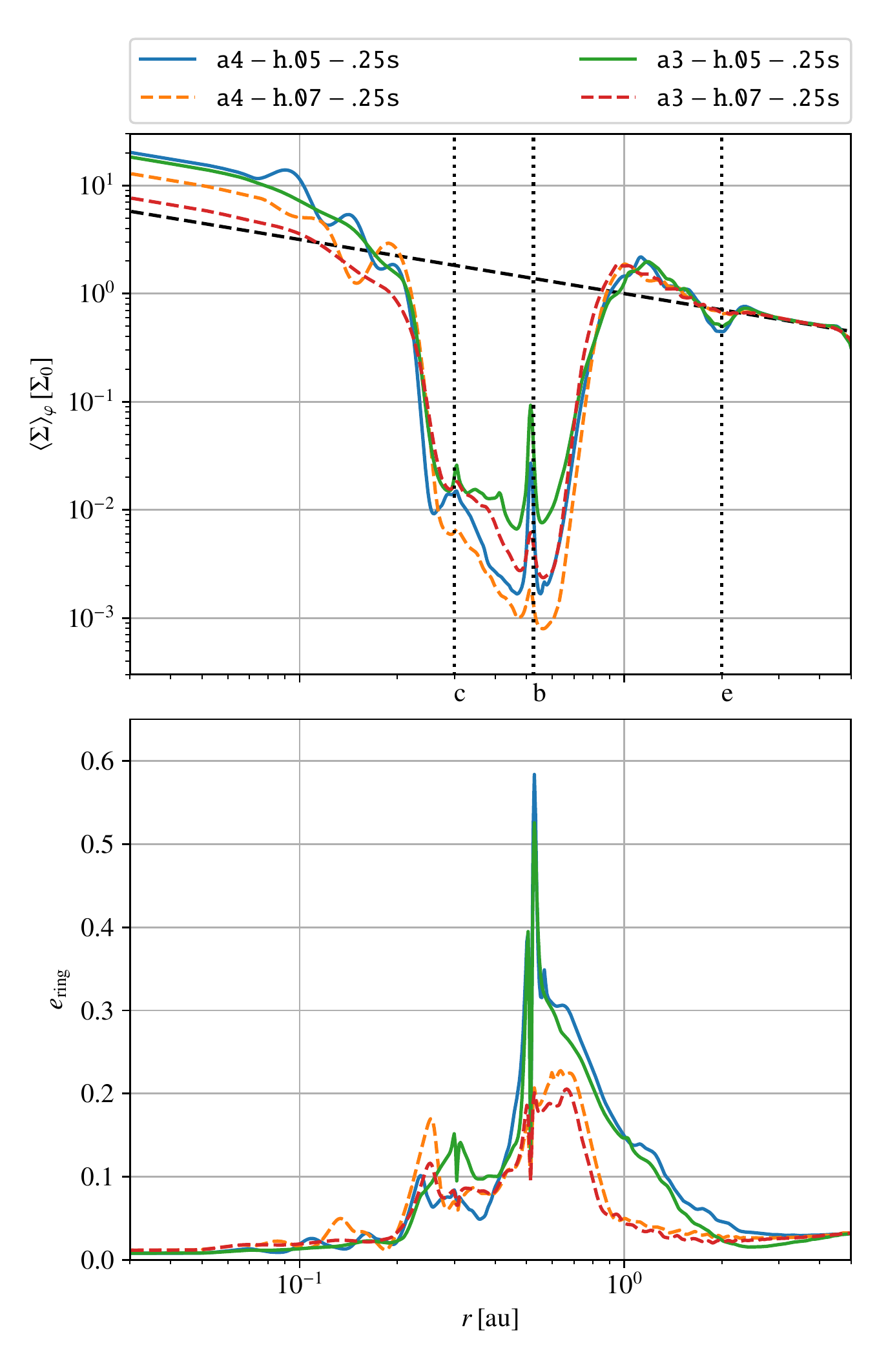}
   \vspace*{-2em}
      \caption{\textit{Top:} Azimuthally averaged surface density profile after the relaxation process for four selected models. The dashed
      black line shows the initial power-law distribution and the three vertical dotted lines mark the positions of the planets.
      Clearly visible is
      the joint gap that the giant planets create.
      While for $h=0.05$ (blue, green) a density depression at the Neptune's location is visible, this is
      not the case for the thicker disk with $h=0.07$ (orange, red).
      The effect of viscosity in smoothing out density waves in the inner disk is visible
      as the difference between the blue and green line in the inner disk region.
      \textit{Bottom:} Eccentricity of rings in the same state for these models. We find that a larger scale-height leads to a lower
      initial eccentricity of the gap edge (blue vs. orange, green vs. red),
      while higher viscosity leads to slightly lower values (blue vs. green, orange vs. red).
              }
         \label{fig:sigrel}
\end{figure}
\subsection{Disk aspect ratio}

\begin{figure*}[h!]
   \resizebox{\hsize}{!}
            {               
   \includegraphics[width=.5\textwidth]{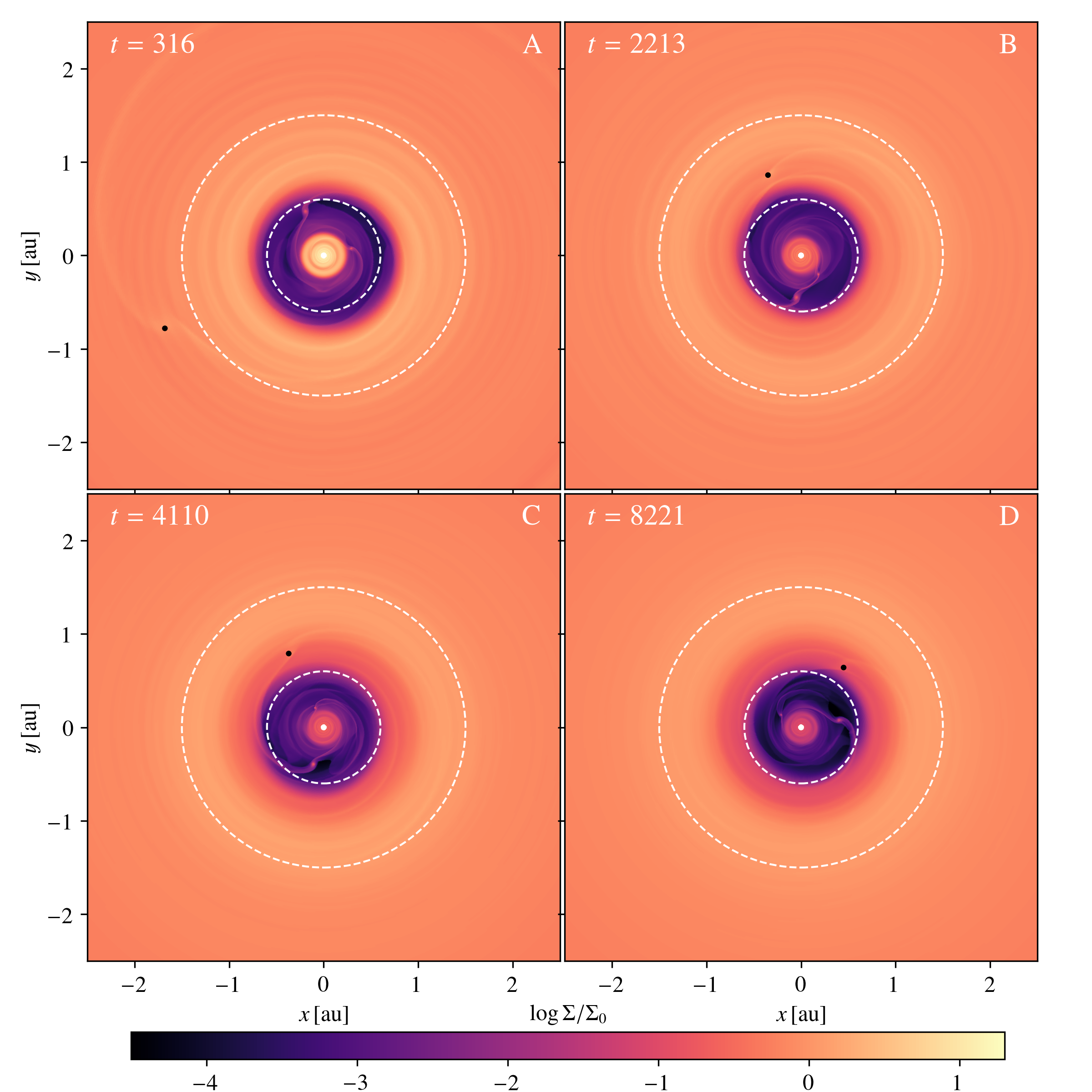}
            }
      \caption{Two-dimensional surface density profiles for the \texttt{a4-h.07-.25s} model.
      The dashed white circles are again of radius $r=0.6 \, \mathrm{au}$ and $r=1.5 \, \mathrm{au}$, respectively. In this model, the
      gap edge more closely follows the inner circle. However, the gap still shows eccentric behaviour.
      At the location of the outer planet, which we here have marked with a black dot, spiral waves driven by the giant planets are clearly
      visible.
              }
         \label{fig:sig2drel-a4h0725s}
\end{figure*}
\begin{figure*}[h!]
   \resizebox{\hsize}{!}
            {                
            \includegraphics[]{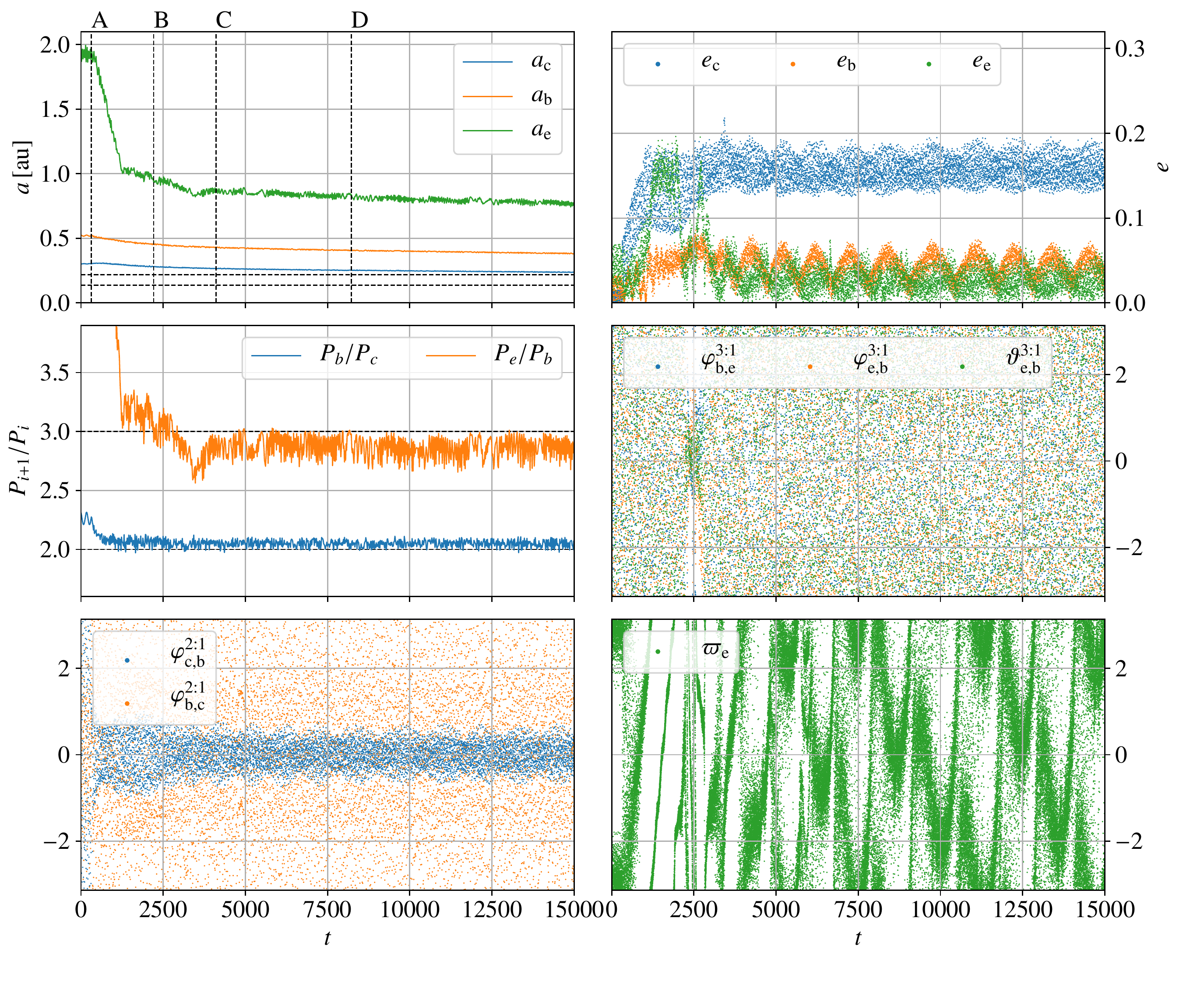} 
            }
   \vspace*{-2em}
   \caption{Similar to Fig. \ref{fig:MigRef} for a simulation with a disk with an increased aspect ratio $h = 0.07$, labelled \texttt{a4-h.07-.25s}. Here we plot the critical angles of the 3:1 MMR of the outer planet pair in the middle right panel and instead of the Laplace angle we
   plot the longitude of periastron of the outer planet in the bottom right panel.
            }
   \label{fig:Miga4h0725s}
\end{figure*}
\begin{figure}
   \centering
      \includegraphics[width=.49\textwidth]{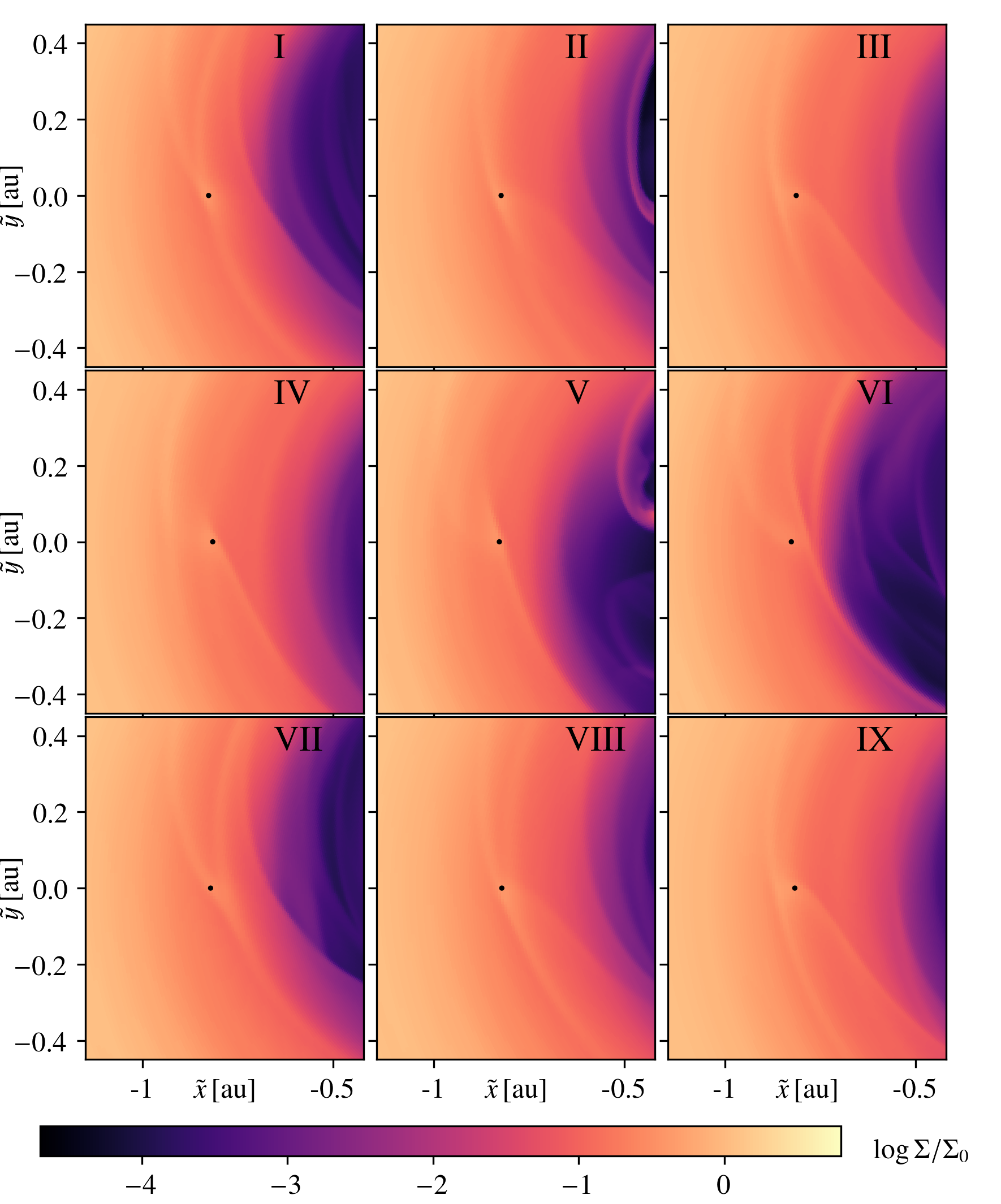}
      \includegraphics[width=.49\textwidth]{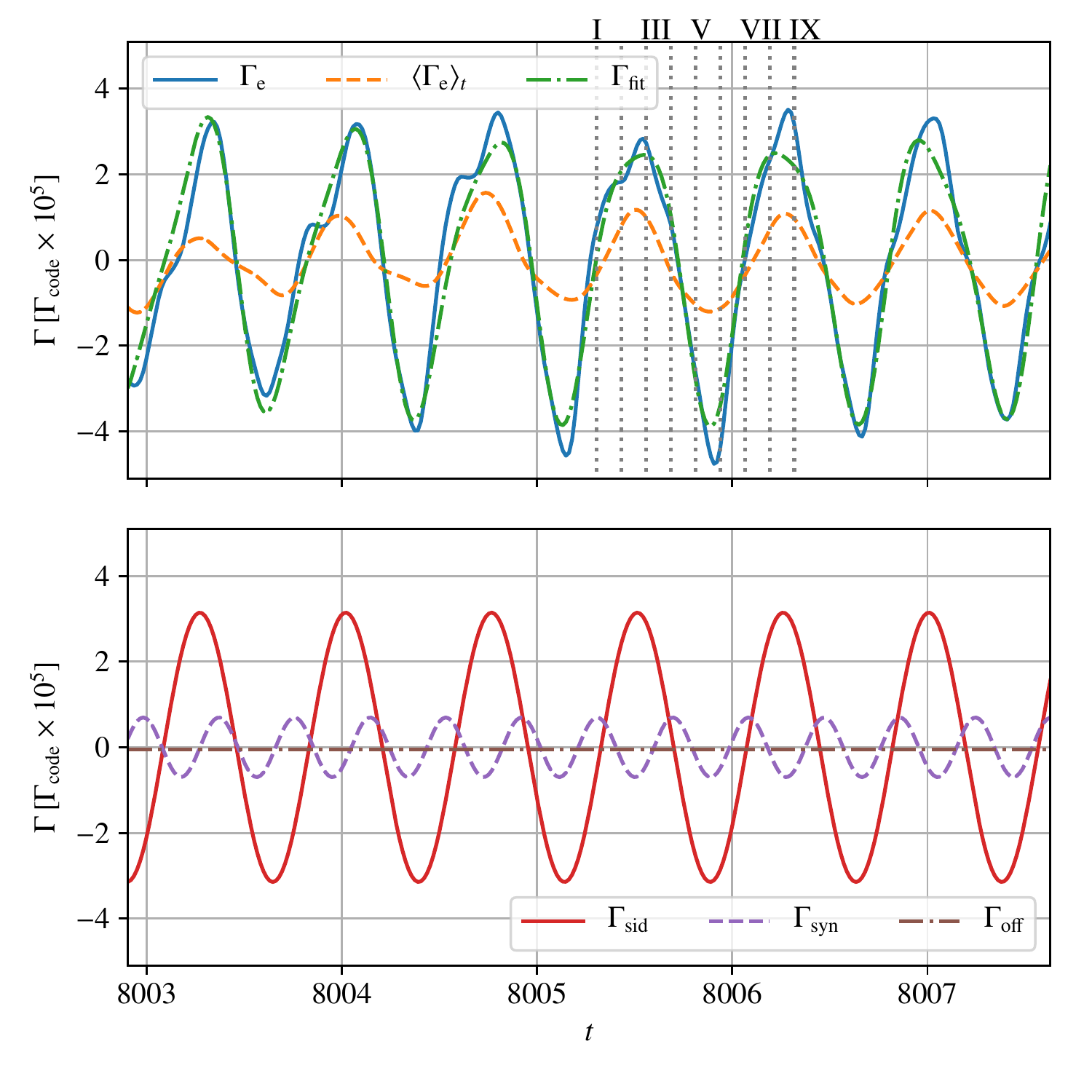}
    \vspace*{-2em}
      \caption{As Fig. \ref{fig:migin} for model \texttt{a4-h.07-.25s}. In this model, migration of the Neptune
      is stalled at around $t \simeq 8000$. In this state featuring a more eccentric disk, the torque
   variation on the sidereal period clearly dominates over that on the synodic period. The torque offset $\Gamma_\mathrm{off}$ is close to zero.
   We note that at the times of a torque maximum, (e.g. III), the outer spiral arm is not trailing the planet but excited at a location leading
   its orbit, which might lead to the positive torque.
              }
        \label{fig:steady}
\end{figure}

Changing the aspect ratio $h$ effectively modifies the temperature which will have an effect on the propagation of spiral waves
and the location of shocks. By changing the pressure, it is also influencing the width and depth of planet-induced gaps.
In a hotter disk, higher planet masses are required to create substantial surface density depletions.

In Fig. \ref{fig:sigrel} we show the azimuthally averaged surface density for several simulations, including the reference model, after the
relaxation process.
A comparison of the blue and green ($h=0.05$) with the orange and red ($h=0.07$)
lines in the top panel of Fig. \ref{fig:sigrel} shows that in contrast to our
reference model (see also Fig. \ref{fig:sig2dRef} A), the outer low-mass planet fails to open a noticeable gap in this case.
The outer edge of the giants' mutual gap is shifted slightly inward.
Regarding the inner edge of the gap, we see less of an increase in the surface density
towards the inner disk for larger $h$, which can be attributed to the higher viscous outflow velocity, that is due to the dependency of an
$\alpha$-type viscosity on $\cs$ and $H$.
In the bottom panel of Fig. \ref{fig:sigrel} the disk eccentricity in the relaxed state is shown. We find that in models
with larger $h$ significantly lower eccentricity is excited in the region connecting the outer gap edge and the
outermost planet.
We plot the time evolution of the surface density for a model with an increased aspect ratio of $h = 0.07$ in the middle panel of
Fig. \ref{fig:sig2dT} as an azimuthally averaged radial profile and show the two-dimensional distribution in Fig. \ref{fig:sig2drel-a4h0725s}.
In the latter, we recognize that the edge of the joint gap cleared by the giant planets more closely follows the guiding circle,
displaying visibly lower eccentricity compared to our reference model.

The dynamical evolution of the orbital elements for the model \texttt{a4-h.07-.25s} is displayed in Fig. \ref{fig:Miga4h0725s}.
For this second-order resonance we plot the angles $\varphi^{3:1}_{\mathrm{e},\mathrm{b} / \mathrm{e},\mathrm{b}}
\equiv 3 \lambda_\mathrm{e} - \lambda_\mathrm{b} - 2\varpi_\mathrm{b/e} $, as well as the angle
$\vartheta^{3:1}_{\mathrm{e},\mathrm{b}}
\equiv 3 \lambda_\mathrm{e} - \lambda_\mathrm{b} - \varpi_\mathrm{e} - \varpi_\mathrm{b}$, related to the mixed $e e'$ resonant terms.
We note here that for a given resonance, either none, one or all of the resonant angles librate, since not more than two
of them can be linearly independent \citep{PapNelson2002}.
Regarding the migration of the Neptune this model shows a similar reduction of its inward migration as it approaches the 3:1 MMR with planet b,
together with a similar increase in its eccentricity. After passing through the 3:1 MMR with planet
b at around $t \simeq 2500$, the Neptune moves slightly inward, until its migration is stalled at around 0.8\,au. From this
point onward, the outer planet pair remains in a non-resonant state.
Around the time of this quasi-steady state, we show a time-series of the surface density and the disk torque acting on the
Neptune in Fig. \ref{fig:steady}. In the top panel, the surface density is again shown in a coordinate frame that is rotated
such that the Neptune is always located around $\tilde{y} = 0$ with $\tilde{x} < 0$.
Regarding the outer planet, we again see how it propagates through the spiral waves excited by the giant
planets and the eccentric gap edge. In this model, however, the Neptune's imprint on the disk is not as pronounced: it is not able to carve
a partial gap into the gap edge and as more gas is available close to the giants, their perturbations pass through its location more intact.
At the time III, close to when we find a torque maximum, the Neptune is located further away from the gap edge than for example at VI.
Regarding the spiral arms around the Neptune, we again find that at torque maximum (e.g. IX), the outer wake slightly leads the planet,
while at the torque minimum at (V), the inner spiral is slightly trailing it.
This picture is reminiscent of the surface density maps for eccentric planets in circular disks, shown for example in Fig. 10 of
\citet{BitschKley2010}.

Regarding the alignment of the eccentric disk, we additionally calculate its longitude of periastron by integrating radially over the surface
density over a selected radial range as a function of $\varphi$:

\begin{equation}
	C(\varphi) = \int\limits_{r_0}^{r_1} \Sigma(r', \varphi) \, r' \, \de r'
	\label{eq:apdisk}
\end{equation}
and we integrate from $r_0 = 0.5\, \mathrm{au}$ to $r_1 = 0.75\, \mathrm{au}$, since we are interested in the alignment of the eccentric gap edge that
interacts with the Neptune.
We then select the minimum of $C(\varphi)$ as an approximation of the argument of apocentre of the disk, since it is less sensitive
to disturbances by the spiral arms. From this we find that from $t \simeq 3000$ onward, the disk precesses in a prograde fashion on a period of
$P_\mathrm{d} \sim 2000 \, T_0$. Comparing this alignment to that of the outer planets orbit $\varpi_\mathrm{e}$ in the bottom right panel
of Fig. \ref{fig:Miga4h0725s}, it is clear that the outer planet's orbit is not locked and co-precessing with the disk. Rather, it shows
prograde precession of varying rates.

The torques and a fit according to \eqref{eq:fittrq} are plotted in the middle and bottom panels.
Again, we are able to recover the synodic and sidereal period, as was the case in the reference model.
This time however, the total torque variations are more than three times larger.
We also find that in this state the variations associated with the eccentric disk, have an amplitude of $\hat{\Gamma}_\mathrm{sid} = 3.15 \cdot 10^{-5}$
, which clearly dominates the smaller contribution of the spiral wake for which we find $\hat{\Gamma}_\mathrm{syn} = 6.97 \cdot 10^{-6}$, which
is consistent with the higher eccentricity of the disk, compared to the reference model.
Over long time-scales the period-ratio of the outer planet pair fluctuates around a steady value, indicating that the Neptune's
migration is dictated by the evolution of the inner giant pair. This finding is consistent with the insignificantly small torque 
offset $\Gamma_\mathrm{off} = -5 \cdot 10^{-7}$ that our fit provides.

None of the models featuring a hot disk display the previously described drastic reversal of the Neptune's migration.
We discuss this finding further in Section \ref{sec:inrap}.

\subsection{Viscosity}
\begin{figure*}
   \resizebox{\hsize}{!}
            {               
            \includegraphics[]{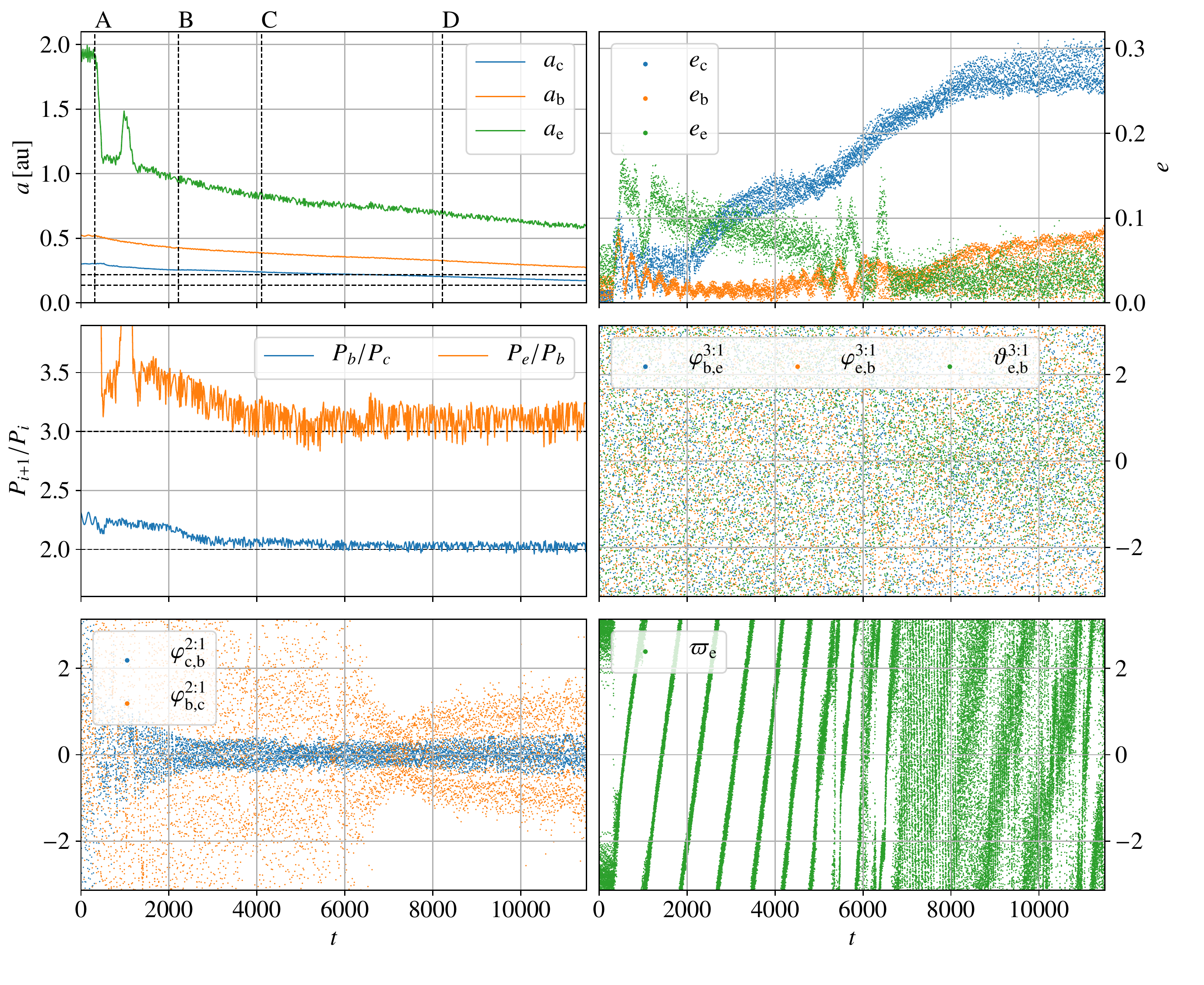}
            }
   \vspace*{-2em}
   \caption{As Fig. \ref{fig:Miga4h0725s}, but showing results for a higher viscosity simulation \texttt{a3-h.05-.25s}.
   }
   \label{fig:Miga3h0525s}
\end{figure*}

\begin{figure}
   \centering
      \includegraphics[width=.49\textwidth]{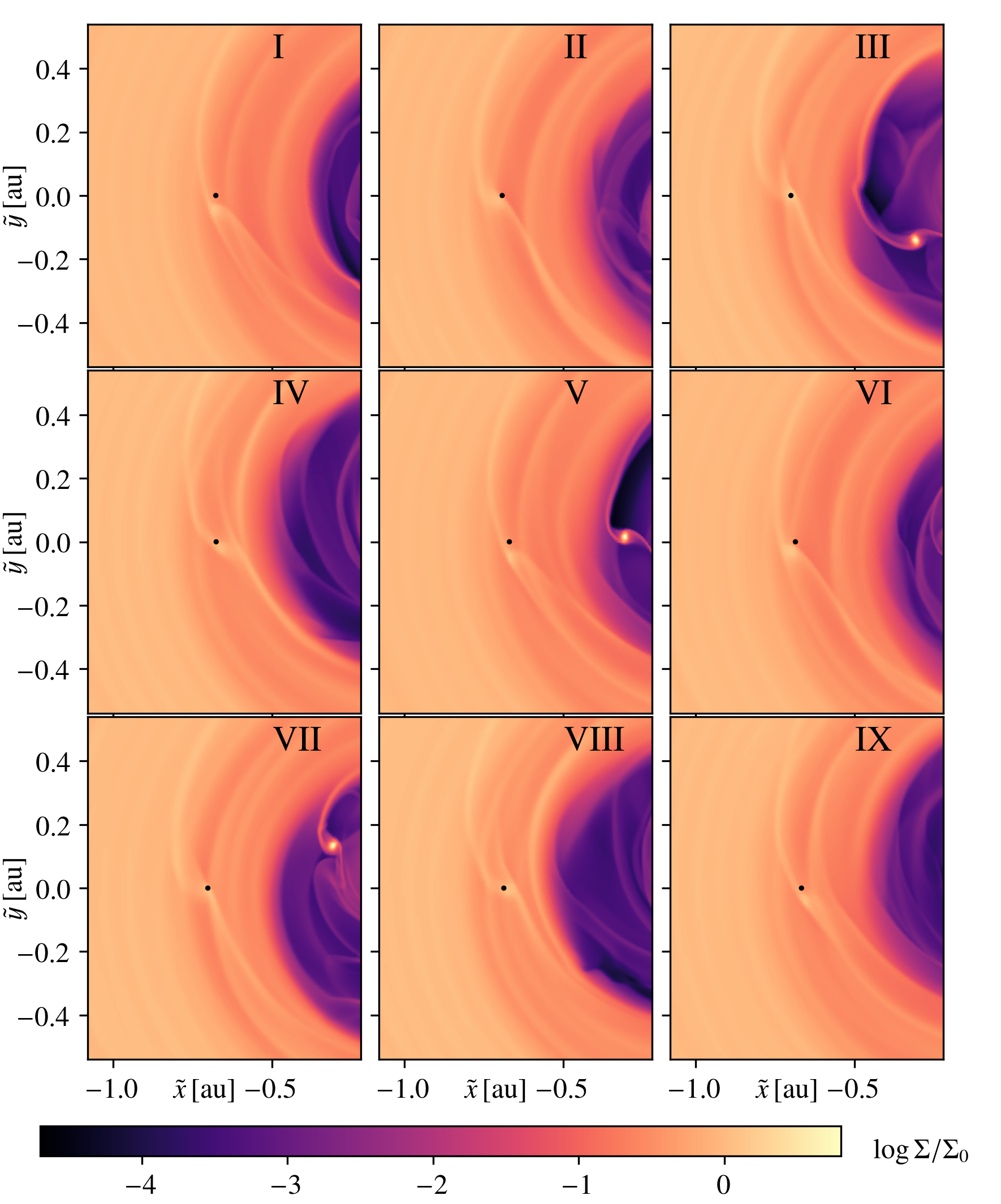}
  \includegraphics[width=.49\textwidth]{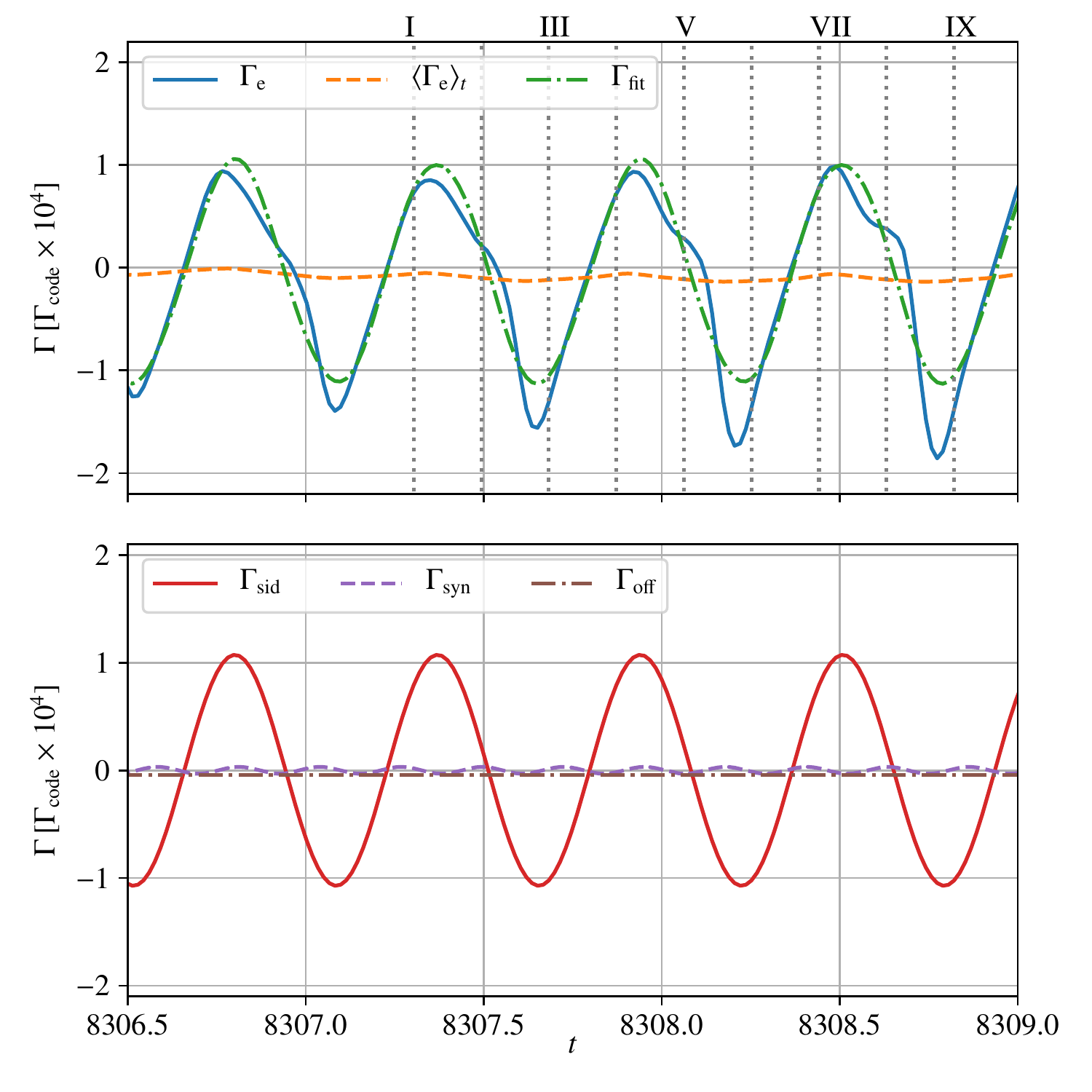}
      \caption{As Fig. \ref{fig:migin} for the higher viscosity model \texttt{a3-h.05-.25s}. In this model, migration of the Neptune
      is stalled close to a period ratio of 3:1, but not trapped in resonance. 
              }
         \label{fig:steadya3h05}
\end{figure}

Another parameter that we varied is the gas viscosity, which tends to close gaps and smooth out features such as spiral waves. These effects
can be recognized in Fig. \ref{fig:sigrel}, by comparing the initial relaxed state for $\alpha = 10^{-3}$ (green, red) and $\alpha = 10^{-4}$
(blue, orange). In the inner disk, the lower viscosity model shows an oscillating surface density up to the radius where our smoothing
prescription sets in and it clearly allows for a deeper mutual gap between the giant planets. We note however, that the partial gap of the outer
planet is hardly affected. The initial disk eccentricity shows only a very slight decrease when increasing the viscosity for a fixed value of
$h$.

In terms of the dynamical evolution of the system, which is shown in Fig. \ref{fig:Miga3h0525s},
we observe mainly the migration rate of the giant pair to be increased.
If there was a very clear and deep gap, the Type II migration of the giant planets would be tied to the viscous evolution time-scale of the
disk. However, in this case with another planet approaching the gap and two giants inside it, the situation is more complex and we
readily observe streams crossing the gap edges.
In any case, viscosity acts to increase the amount of gas located close to the massive planets and enhances their migration rates.
As a consequence of this continued migration at a higher rate, the eccentricities of the giant planets keep rising
above their observed values, indicating that sufficient eccentricity damping is not provided in this case.
The outer planet's migration rate does not change substantially during the initial migration phase, which is still rapidly inward.
As in the reference model, we observe a phase of divergent migration, which is then followed by the Neptune approaching the 3:1 period ratio
where it is stopped. We note however, that it is not trapped in resonance there, as none of the critical angles librate.
The surface density distribution is displayed in Fig. \ref{fig:steadya3h05} and shows that the disk remains eccentric in this
model.
Investigating the torques acting on the Neptune in this situation, we find that the amplitude of the oscillations is roughly an order of
magnitude larger compared to  our reference model. In this model, this amplitude is subject to change on time-scales of tens of orbits
of the Neptune. This is most likely related to the alignment of the planet's orbit and the disk, which changes rapidly in the later stages
of evolution, since the planet's orbit precesses at a much higher rate than the eccentric disk.
We note that in this more compact system, the disk precesses at an increased rate, compared to the previously discussed cases.
While we still recognize the sidereal period in the signal, this changing amplitude makes our fit deviate
on short time-scales. We also note, that we are not able to recover a meaningful signal on the synodic period this time.
These strong torque oscillations keep the planet from settling into the second order resonance, as it is constantly experiencing strong kicks.
The Neptune remains stalled in this situation up to the point where the giant planets migrate past their present locations, where we stop the
simulation.

When testing models with an even higher viscosity of ${\alpha = 10^{-2}}$, we found that they allow the giant planets to migrate at such a high
rate, that the Neptune is unable to approach them, before the giant pair migrates past their observed location.

\subsection{Disk mass}

Thinking of single planets, one could naively expect that varying the disk mass by adjusting $\Sigma_0$ would lead to a similar evolution of the
system and only affect the time-scale on which it occurs, given the locally isothermal nature of the disk.
Such a similarity of the models can be found in the surface density profiles of the relaxed initial state of the system that we recover,
which is virtually the same for four different disk masses and simply scales with $\Sigma_0$.
Since the planets are forced to stay on circular fixed orbits, the relaxed disk eccentricity profiles are also identical,
which can be understood since the eccentric modes of the disk are expected to be independent of $\Sigma_0$, when
the self-gravity of the disk is neglected \citep{Teyssandier2016}.
By adjusting the disk mass and leaving the planet masses untouched, the ratio of angular
momentum stored in the disk over that of the planets $L_\mathrm{disk}/L_\mathrm{pl} \sim \Sigma_0$
in this state, is changed in a linear fashion.
As soon as the planets are released it becomes evident that scaling the surface density does not affect the migration rates of all
planets in the same way.

In the \texttt{a4-h.05} simulation, where the viscosity and scale-height are identical to our reference model, but the surface density is
multiplied by a factor of four, the initial migration of all planets is more rapid which has an effect on the inner planet pair, leading to a resonant state, where initially only one resonant angle librates with a large amplitude of
$\sim \pi/2$, while the other remains in circulation.
The eccentricity of the innermost planet c shows large-amplitude oscillations between $\sim 0 - 0.2$. During this phase the outer planet repeats multiple phases of in- and outward migration. At roughly $t \sim 6000$, the giant pair
switches to a state where both angles librate with amplitudes of about unity. The outer planet then crosses the 3:1 MMR, and stops migrating at around $0.6 \,\mathrm{au}$, where it remains for the rest of the simulation, while the giant pair slowly
migrates inward. While in this simulation the planet was able to clear a partial gap as well, it was unable to circularize
the disk in this model, as opposed to the reference case.
This can be interpreted of a consequence of the larger angular momentum deficit in the disk that is due to the increased
$\Sigma_0$. Since the planet mass is the same, it may not be able to provide sufficient eccentricity damping.

Varying the disk mass can also have more subtle effects, like modifying the apsidal precession rate of the gap and planets $\dot{\varpi}$
\citep{Kley2005}.
For more massive disks, self-gravity becomes important and is expected to contribute to a prograde precession of the disk.
We generally observe the phases of rapid migration to be more pronounced and in higher disk mass models as well as
stretching over larger radial ranges.
In a simulation with a disk mass that is half of the reference value, the migration rate of all planets is reduced as expected, but leads to the
successful formation of the Laplace resonance as well, which suggest that it could have occurred at a late stage of disk evolution.

\subsection{Initial position of the planets}
Another free parameter is the initial position of the planets, which we explore by repeating our reference simulations
having the planets start at different locations.
First we consider a case where we place the Neptune further in at $a_\mathrm{e,init} = 1.5\, \mathrm{au}$ or further out at $a_\mathrm{e,init} = 3.0\, \mathrm{au}$.
In the former case, the initial migration phase behaves similarly, but planet e repeats the reversal of its migration around the 3:1 period
ratio twice.
As it later crosses the 3:1 period ratio, the innermost planet shows strong short-period oscillations in its eccentricity, resulting in a large
amplitude libration of the resonant angle of the giant pair. As the Neptune approaches the 4:2:1 MMR, this time it is scattered outward to
beyond 2\,au, where it restarts the journey inward. In the model where we introduce the Neptune at 3\,au, the behaviour is
almost identical to the reference model. We find that the early phase of divergent migration reverses again at a similar
radius, indicating that it is related to the density maximum at the gap edge, rather than the partial gap that is created during the relaxation
process, which is located further out in this simulation.
After this stage, the evolution of the system is virtually identical to our reference model and the same Laplace resonance is formed with
very similar libration widths.
This indicates that in our reference model we started the outer planet far enough outside to be compatible with an origin from even
further out in the disk.

In another simulation, we started the inner planet pair at the same period ratio, but move it further in to
$a_\mathrm{c,init} = 0.2\, \mathrm{au}$ and $a_\mathrm{b,init} = 0.35\, \mathrm{au}$, respectively.
In this model, the giant planets again quickly engage in a 2:1 MMR with both angles librating after $t \simeq 2000$,
as the innermost planet settles to a similar eccentricity of $\simeq 0.21$. We initially observe
convergent migration of the outer planet pair, which is then stalled close to a period ratio of 5:2, outside of resonance
for several thousands of orbits. In this case, the disk maintains higher eccentricity is than in our reference model, which appears to be
related to the orbital location of the giant planets.

\section{Discussion}
\label{sec:discussion}
In this section we discuss our results and compare our findings to previous works.

\subsection{Disk eccentricity evolution}
Comparing the disk eccentricity profiles after introducing the planets and letting the disk adapt for several physical parameters of the
disk, we found that in a colder disk ($h = 0.05$) the region between the gap and the outer planet reaches substantially higher
deviations from circular motion than in hotter disks ($h=0.07$). Viscosity on the other hand only slightly changes this initial profile,
with lower values leading to higher eccentricities.
This can be understood as an effect of the different gap profiles in the hotter and colder disk. In the two cases, Lindblad and corotation
resonances inside the gap, that can act to damp or increase the disk eccentricity \citep{GoldreichSari2003,Ogilvie2003,Teyssandier2016} can be of different significance,
simply due to the varying availability of gas to communicate the disturbances. These general trends regarding dependence on the disk
scale-height and viscosity that we find here are consistent with those described by \citet{KleyDirksen2006}, who considered a single fixed,
gap-opening planet on a circular orbit, and found that in their model mainly the 1:3 outer Lindblad resonance is causing the disk eccentricity
to rise.

While the above described trends are clear in the state where the planets are still fixed and orbiting circularly,
the time evolution of the disk and planet eccentricities are much more complicated and strongly interconnected.
In cases where the Neptune opens a partial gap and is able to approach the inner planets, as for example in our
reference model, we find that as a consequence the disk eccentricity is reduced. We expect this to be due to a combination
of weakening the eccentricity-exciting resonances of the giant planets by depleting the gap and possibly 
eccentricity-damping provided by the Neptune's own resonances. The ability of the outer planet to circularize the disk is expected to depend
on its mass compared to the disk mass.
Recently, \citet{Thun2018} studied the migration of planets in circumbinary disks and found a trend for more massive planets to
shrink and circularize initially eccentric gaps created by the binary. To investigate this, we performed a single simulation with reference
disk parameters but increased the outer planets mass to $M_\mathrm{e} = 3 \cdot 10^{-4} M_\ast$. In this simulation, the outer planet
migrates only inward and quickly renders the disk circular, approaching the 2:1 MMR, which confirms this trend.

\subsection{Initial phases of rapid migration}
\label{sec:inrap}
Especially for parameters that allow the Neptune to open a partial gap, creating a significant mass-deficit in its corotation region, we found
initial phases of fast inward migration, which is then followed by a phase of rapid outward migration, that is started as the eccentricity
of the planet increases close to a period ratio of 3:1 with planet b.
Once it begins its outward journey, the Neptune reduces its eccentricity again.
This mode of migration shows dynamical features: while the density profile is not changing significantly between subsequent
phases of in- and outward migration, the planet is able to rapidly move through this region in both radial directions, depending on its
initial radial velocity that sets off this phase of rapid migration.
Simulations featuring an aspect ratio of $h=0.07$ do not show this behaviour.
This influence of the disk scale-height can be understood directly within the framework of Type III migration
\citep{Masset2003,Artymowicz2004}.
Since in this case, the Neptune is unable to create a partial gap, there is no sufficient co-orbital mass deficit to make it susceptible to this
type of dynamical migration which provides a positive feedback on migration.
In models that involve higher disk masses, we find that these phases of rapid in- and outward migration repeat more often, showing higher
migration rates and extending over larger radial ranges.
The location where this occurs is close to the density maximum outside the gap, which can provide a strong density and vortensity gradient
which makes for the large migration rates.

In models where we find a temporary migration reversal, this always occurs close to the 3:1 MMR where the Neptune's eccentricity is excited.
However, identifying the mechanism that sets off the temporary rapid outward migration requires further investigation, especially
since this did not occur in a simulation with higher resolution, despite similar values of the Neptune's eccentricity
(see Appendix \ref{sec:AppRes}). Since the overall evolution of the system turns out very similar, we are confident that this discrepancy will
not change our main conclusions.

\subsection{Stalling the Neptune's migration}
Many of our models show the Neptune being unable to approach the observed 2:1 resonance as its migration is halted when it approaches the gap.
This qualitative behaviour resembles the results obtained by \citep{PG2012} and \citet{Baruteau2013} who also considered the migration of a low-
mass planet orbiting exterior to a giant planet and attributed this behaviour to the interaction of the outer planet with wakes driven by its
interior companion. In our case, the situation is more complicated due to the eccentric nature of the disk, which introduces an additional
variation of the torque acting on the Neptune.

Using a simple fit, we were able to match the large amplitude variation of the torque and, in steady states, recover the sidereal period
of the outer planet and the synodic period of the outer planet pair. While this simple picture allows us to understand the origin and
variability of the
torque, the quantity that sets the long-term direction of migration is the average torque.
As pointed out by \citet{Cresswell2007}, for eccentric planets, both the torque and the power (change in angular momentum and
energy) due to disk forces determine the changes in $e$ and $a$. This can lead to situations, where a positive torque but a negative power
are observed, leading to inward migration, which would not be expected from the torque alone. Since the eccentricities are usually
low in states where we display torques, we restrict ourselves to discussing them. The migration is also directly evident from the evolution
of $a$ and not inferred from the torque or power in our case.

As explained by \citet{Papaloizou2000}, who considered eccentric low-mass planets in circular disks with orbital eccentricities of order
$e \simeq h$, the varying orbital velocity of the planet can lead to a shift of Lindblad and corotation resonances and situations where
the outer (inner) disk exerts positive (negative) torques.
In many of our models however, the disk gas close to the Neptunes location is on orbits of even higher eccentricity, which provides
additional variations of the relative orbital velocity of planet and disk.
This may contribute in the same way to the shift of resonances and in consequence modify the torque, which has been described by
\citet{Papaloizou2002}, considering again low-mass planets, this time orbiting in an eccentric disk.

However, we observe that in models where the disk remains eccentric, the spiral wakes
excited at the resonances close to the Neptune shift their locations, similar to what was found for example by
\citet{BitschKley2010}, who considered eccentric planets in a circular disk.
When these torque variations that we associate with the eccentricity of the disk are large enough, they might provide a possibility
for the planet to cancel other torque contributions on average, if they are not strictly symmetric in time.
Such an argument has been made by \citet{PierensNelson2008}, who considered a single planet migrating in an eccentric circumbinary
disk. Observing similar torque oscillations related to the disk-planet alignment, they argued that for an eccentric planet, the varying orbital
velocity can lead to such an asymmetry, as it spends more time at apocentre, where the planet experiences a torque maximum in their models.
Comparing the occurrence of torque extrema to the orbital radius of the planet, we find no clear trend for maxima (minima) occur close to passage
of pericenter (apocenter) of the planetary orbit. This is probably related to the fact that the planetary orbit is not locked
to the alignment of the disk, but precesses on a shorter time-scale.

If the outer planet is massive enough to circularize the disk and reduce the amount of gas inside the giants' gap, this stopping mechanism can be
overcome, since both effects will reduce the amplitude of the torque variation and allow the outer, typically negative Lindblad torque to
dominate, leading to inward migration.
This picture is consistent with our finding that in all cases where the torque is clearly dominated by variations due to the eccentric disk,
the Neptune's migration stops outside of resonance at the gap edge. We also found that by decreasing the mass flux into the gap, the
perturbations caused by the giants are weakened due to the lack of gas to mediate them, and the outer planet is able to approach and finally
settle into the 2:1 MMR.

In order to separate planet-planet and planet-disk interactions, we followed the procedure used in \citet{PG2012} and \citet{Baruteau2013} and
repeated our simulation after switching off the gravitational interaction of the Neptune with the other planets in a dedicated simulation
(labelled \texttt{NG}). For this model, we find a similar decrease in its migration rate and increase in eccentricity, as it approaches a 3:1
period ratio, followed by a phase of outward migration. This clearly shows, that the interaction of the outer planet with the perturbed disk
alone can cause this behaviour. In this simulation the outer planet repeats its in- and outward migration several times, as the 3:1 MMR is not
available to trap it. However, after several phases of convergent and divergent migration, it later on settles close to
a 3:1 period ratio in a non-resonant state. This situation is similar to the findings of \citet{Baruteau2013}, who noted that although direct
planet-planet interactions are switched off, there might still be weak resonant coupling due circumplanetary gas accumulation and the spiral
wake. The inner planet pairs behaves slightly different to the reference model, as they keep migrating more quickly and the eccentricity of
planet c remains lower, only one of the resonant angles is librating.

We reproduced this stalling of migration in simpler, higher resolution simulations that featured only one fixed giant planet,
that is again able to excite substantial eccentricity in the disk, and the Neptune orbiting exterior to it.
Results from these models confirm that the Neptune's inward migration can be halted
outside the resonance, when the density close to the giant remains higher due to a different treatment of the BC.
If some viscous evolution of the disk and accretion through the inner boundary is allowed however,
the Neptune approaches the giant and is able to reduce the disk eccentricity, which confirms the trend which we saw in the three-planet
models.

\subsection{Comparison to previous works}
\label{sec:compprev}

Regarding the migration of the inner planets, we agree with the general finding of \citet{Crida2008}, that an inner disk can provide
eccentricity damping at a sufficient rate to the innermost planet in the resonant chain. Different to our study, they were however only
interested in the giant pair and considered higher viscosity $\alpha = 10^{-2}$ and only the case of $h=0.07$, which results in typically
larger migration rates compared to our parameter set.
Additionally, by varying the treatment of the inner boundary condition, we found that it can affect the eccentricity damping
time-scale of the inner planet and in extension the dynamical evolution of the system as a whole.
Our treatment of the inner disk is similar to that in \citet{Crida2008}, where its innermost part evolves viscously. We also note that our
simulations cover a much longer time, and the planets typically migrate over larger radial ranges.

Investigating the interaction of the outer planet pair, \citet{PG2012} conducted a similar two-dimensional, locally isothermal simulation with
planet masses according to GJ 876 b and e, however keeping the giant planet fixed and switching off the  gravitational planet-planet
interaction.
For parameters similar to our setup they found for $h = 0.05$, that the Neptune stalls outside
resonance. In our two-planet simulations where we keep the giant planet fixed, like they did, we find very similar behaviour.
While the authors of that study attributed the stalling purely to planet-wake interactions, we find that the eccentricity of the disk
plays an essential role.
After artificially decreasing the surface density in their models, they were able to get migration into the 2:1 MMR.
In our model, where the surface density self-consistently drains onto the star, this leads to the successful formation of the resonance.

In many ways the situation that we encounter for the planets in this particular system is reminiscent of the migration of a planet
that orbits exterior to a central gap or cavity created by a stellar binary. Investigating such a situation, \citet{PierensNelson2007} found
that low-mass planets can be trapped at the edge of the inner cavity and accounted this to a cancellation of a positive corotation torque, that
is due to a strong positive surface density gradient, and a negative differential Lindblad torque. In a following work, considering higher
masses for the outer planet, \citet{PierensNelson2008} also found phases of rapid reversed migration that starts when the planet gains some
eccentricity. They also report similar torque oscillations due to the eccentricity of the planet and an eccentric cavity edge.

\section{Summary \& Conclusions}
\label{sec:sum}
In this section we summarize our findings and give an outlook on possible directions of future studies.
Using locally isothermal two-dimensional hydrodynamical simulations, we studied the migration of three planets of masses
corresponding to those discovered around GJ 876. By varying physical disk parameters as well as the numerical modelling
approach, we were able to find cases where the three-body resonance is formed and cases where the outer planet's migration
is stalled outside of resonance. We now summarize the characteristics of models that produced similar results.

\begin{itemize}
\item[-] \textit{Successful formation of the Laplace resonance:}\\
Depending on the disk's viscosity and scale-height, the outer planet can open a partial gap and over long time-scales weaken
the eccentricity-related repulsion by starving the giants' gap.
In these cases the planet is able to reduce the disk eccentricity, possibly through a combination of starving the
eccentricity-exciting resonances of the giant planet and providing eccentricity damping itself.
As a result, associated torque variations are reduced and the outer planet is able to slowly migrate along the gap edge and settle into a 2:1
MMR with the giant planet, completing a 4:2:1 Laplace MMR that is similar to the observed state of the system.
A late migration of the outermost planet into resonance seems likely from our findings, since a low disk-mass model also resulted
in the formation of the Laplace resonance and the effects that lead to divergent evolution are expected to be weaker then.
Disk parameters like a lower aspect ratio and viscosity, that favour the formation of the resonance are also compatible with this scenario.
In a model where the giant planets are started closer to the star, the outer planet is unable to approach them, which
suggests that the resonance around GJ 876 was formed in a wider configuration and became more compact over a possibly long time-scale until the disk
dispersed.
In our simulations, the Laplace resonance is in a state where only the inner planet pair is in ACR, while the outer pair is not.
It remains to be explored whether this can change as the resonant chain becomes more compact until it matches the observed
planet locations.

\item[-] \textit{Stalled migration in an eccentric disk with a cavity:}\\
We find in many models, that the interaction of an outer lower mass planet with the
wake driven by an interior giant planet and an eccentric gap edge can slow down or even halt the outer planet's inward
migration. We showed that torque acting on the outermost planet shows oscillations on characteristic periods of the system
and attribute them to a shift of resonances due to eccentricity of the disk and planets, as well as variations
due to the spiral wakes of the interior giants. In order to disentangle different contributions, future studies should consider simpler
situations and investigate their dependence on physical parameters of the disk and planets.
Whenever the torque variations due to these effects are large, settling into resonance is prevented, even if the
outer planet is stalled close to a commensurability.

\item[-] \textit{Boundary conditions:}\\
Conducting many numerical experiments, we found that the choice of the inner BC
and damping prescriptions used in its vicinity can have strong effects on the dynamical evolution of the system.
Due to the high planetary masses and compactness of the system, we find a damping of the velocities close to the inner
boundary to be necessary to avoid the formation of an artificial cavity, where the density is strongly reduced.
A viscous evolution of the inner disk can still be accounted for by using velocity damping towards viscous outflow in combination
with an outflow-only BC.
\end{itemize}

In our simplified approach, we modelled a two-dimensional, locally isothermal disk of constant aspect ratio. More realistically, the disk
will be affected by irradiation and heating due to viscous dissipation and spiral shocks driven by giant planets, since
not only angular momentum, but also energy is transferred \citep{Rafikov2016}.
As has been shown for single planets already, radiative effects can substantially affect their migration rates, mainly by modifying the
corotation torques \citep{Paardekooper2011}. A more realistic treatment of the disk thermodynamics is of special interest for the migration of
an exterior lower-mass planet, since it might change the disk's eccentricity structure \citep{Teyssandier2016} and the interaction and shape of
the giant planets' wake. In order to study the role of the disk and planet eccentricity in more detail, simpler 
situations should be considered, as for example a single planet migrating in an eccentric disk with a cavity.
In order to find a simple description that could for example be used in N-body simulations, the dependence on the physical parameters of the
disk and planet should be investigated in more detail in future work.

In a three-dimensional model, additional effects could become important. Not only can the the growth-rates of eccentric modes be
modified \citep{Teyssandier2016}, but the disk would be allowed to become warped and second-order inclination resonances would be available to
the planets, which have been considered previously \citep{LeeThommes2009}.

While we considered the planet masses to be fixed at their current values, mass growth by gas accretion adds additional pathways to the
formation of this system. For example, a scenario where the planets grow already trapped in resonance could be imagined. However, given the
potential
starving of the inner planets that we observed, this seems rather unlikely. Due to the high masses of the giants, they must have accreted
substantial amounts of gas during the gas-rich phase of the disk.

The migration of a slightly eccentric planet in an eccentric disk warrants further investigation, since it is also of great
interest for the migration of circumbinary planets.

Since our study has revealed that the migration of the outer planet can be very different from a smooth, purely dissipative
evolution, the observed torque variations might act in a similar way to turbulence in preventing settling into a regular state,
which was proposed by \citet{Batygin2015}.

%__________________________________________________________________

\begin{acknowledgements}
		We thank the anonymous referee for helping to improve the presentation of the paper and to clarify some arguments made.
		NPC thanks Pablo Benítez-Llambay for support with using and modifying the code and helpful discussions.
		RK acknowledges financial support via the Emmy Noether Research Programme funded by the German Research Foundation
		(DFG) under grant no. KU2849/3-1.
		The authors acknowledge support by the High Performance and Cloud Computing Group at the Zentrum f\"ur Datenverarbeitung of the 
		University of T\"ubingen, the state of Baden-W\"urttemberg through bwHPC
		and the German Research Foundation (DFG) through grant no INST 37/935-1 FUGG.
		All of the numerical simulations were performed on the bwForCLuster BinAC,
		supported by the state of Baden-W??rttemberg through bwHPC, and the German Research Foundation (DFG) through grant INST 39/963-1 FUGG.
		\textit{Software:} We acknowledge the use of the following software
		and thank the respective authors for making it publicly available: \texttt{matplotlib} \citep{Hunter2007}, \texttt{Jupyter}
		\citep{Kluyver:2016aa} and \texttt{FARGO3D} \citep{FARGO3D}.
\end{acknowledgements}

%-------------------------------------------------------------------

% for the bibliography, at the end
\bibliographystyle{aa} % style aa.bst
\bibliography{GJ876} % your references Yourfile.bib

\begin{appendix}

\section{Interaction of an outer Neptune with only one fixed giant}
In order to simplify the situation and study only the interaction of the outer planet pair, we conducted simulations that excluded the inner giant
and where we kept the remaining giant on a fixed, circular orbit. For these simulations, we move the inner boundary further out to
$r_\mathrm{inner} = 0.1\,\mathrm{au}$ and use a closed inner boundary with the damping prescription, which we in these models apply also to the
surface density, to damp it to its initial value.
We chose this setting to create a situation more similar to that considered by \citet{PG2012} and remove the additional complexity due to the
migration of the inner planet and the time evolution of the inner disk.
In these simulations we use an increased resolution of $N_r \times N_\varphi = 576 \times 926$ and we compare our results to those of \citet{PG2012} in Section \ref{sec:compprev}.

In a simulation using the reference values of $\alpha = 10^{-4}$ and $h = 0.05$, the outer planet behaves virtually
identical in its initial migration which is again rapid and shows a one-time reversal. After crossing the 3:1 MMR, however, its migration is
stalled at a period ratio of $\sim 2.6$, where it remains in a low-eccentricity orbit with $e < 0.05$ for at least 10000 orbits.
The surface density profile settles into a steady state where the gap is shallower than in our full three-planet model featuring the viscous
outflow BC.
Repeating this exercise with $\alpha = 10^{-3}$ and $h = 0.05$, the outer planet only migrates in a convergent fashion, never reversing, until
it is stalled close to the 3:1 MMR but outside of resonance.
In these stalled models, the precession of the outer planet's orbit becomes locked to the precession of the eccentric disk, such that they
are aligned with little variations.

When we repeat this simpler two-planet model returning to using the viscous outflow BC and reference parameters, the evolution is initially
similar but deviates as the gap and inner disk are drained, as the Neptune stalls the gap edge.
Ultimately, the Neptune passes the 3:1 MMR and migrates closer inward
to briefly become trapped in a 5:2 MMR, from which it escapes then to migrate smoothly up to the 2:1 MMR with the giant planet, where
it remains.
This simple test illustrates the effect that the treatment of the inner boundary can have on the planet-wake interaction and the resulting
orbital state a simulation settles towards.
In this model, the precession of the outer planet shows a trend to follows that of the disk partly but deviates often to precess at a higher
rate.

In these simpler models we are again able to recover the two characteristic periods of the torque variation as the sidereal and
synodic period of the Neptune. Again the amplitude associated with the eccentric disk dominates the contribution of the giants' wake.
We also confirm that in the model featuring the viscous outflow BC, where the planet is able to migrate inward, we find a negative torque offset
and the amplitude $\Gamma_\mathrm{d}$ to be a factor two smaller than in the model where migration halts outside of resonance.

\section{Effect of the damping prescription and the inner BC}
\label{sec:bound}

\begin{figure}
   \centering
   \includegraphics[width=.5\textwidth]{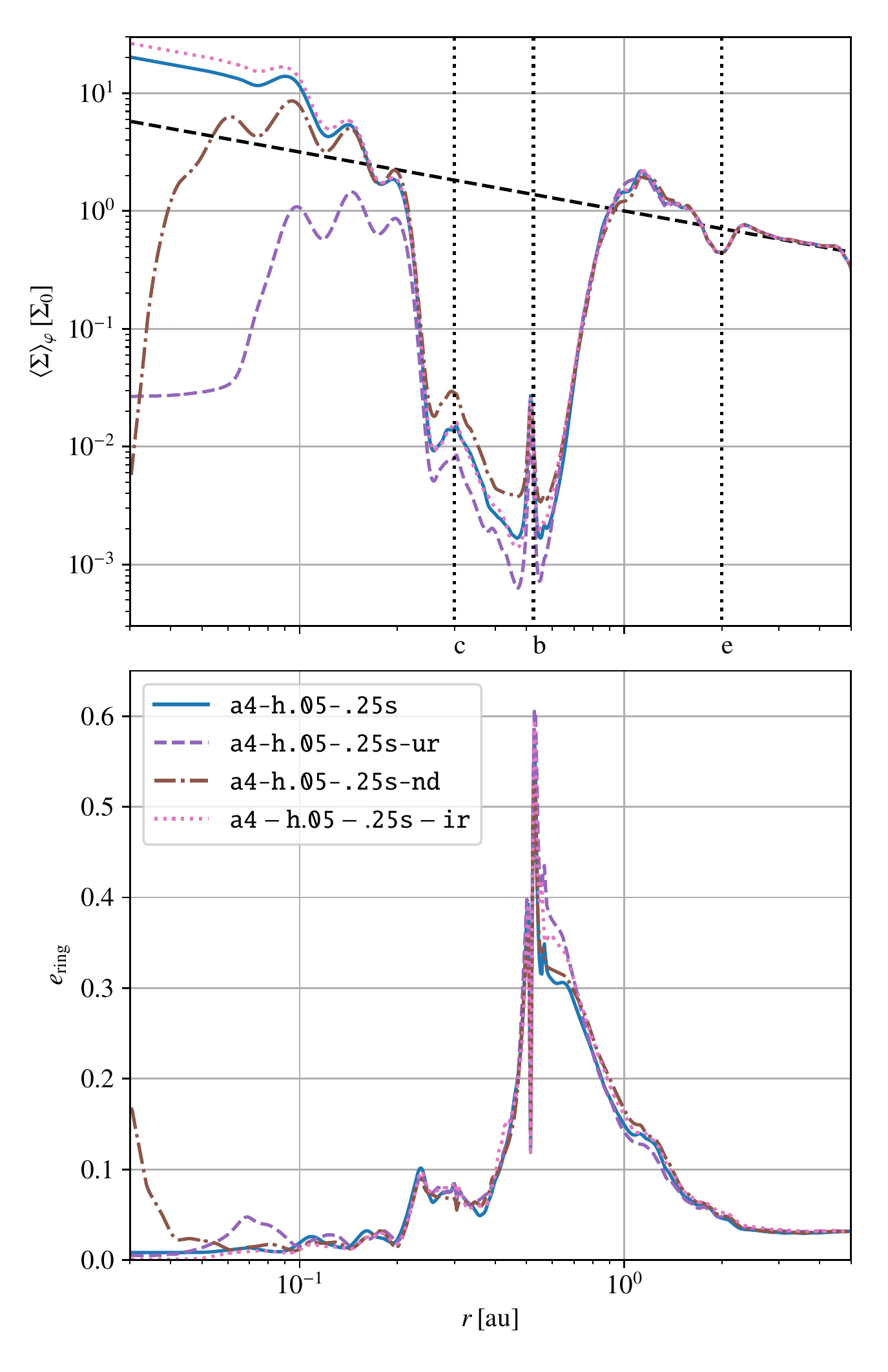}
   \vspace*{-2em}
      \caption{
	 As Fig. \ref{fig:sigrel}.
      Here, we show the relaxed state of a simulation with an outflow
      inner boundary condition without damping the radial velocity, but damping $v_\varphi \rightarrow v_\mathrm{K}$  (\texttt{ur}, brown) and a
      model with no damping at the inner boundary at all (\texttt{nd}, purple). In the former case the artificial cavity is clearly created in
      the damping zone, while the latter, undamped model only shows a surface density decrease at the innermost edge of the domain.
      The pink dotted lines (\texttt{ir}) show results of a model that uses the standard damping but features a closed inner BC. It only
      deviates slightly, hosting more mass in the inner disk.
              }
         \label{fig:sigrel_app}
\end{figure}

We dedicate this section to the discussion of the effect the choice of the inner boundary condition has on the general structure and evolution of the system.

First, we compare models that adopt the same inner radial boundary condition that allows for outflow only but treat the damping differently.
Using this BC, whenever gas sitting close to the inner boundary is moving across it, it is irreversibly lost from the simulation.
This effect becomes evident by comparing the relaxed surface density profile of the reference model (blue) and the simulation with radially
undamped outflow (\texttt{ur}, purple) and model without any damping (\texttt{nd}, brown) in Fig. \ref{fig:sigrel_app}.

In the model labelled \texttt{ur} where we do only apply the damping prescription to $v_\varphi$ but \textit{not to} $v_r$,
an artificial cavity in the inner disk is created already during the phase when the planets are still held on fixed, circular orbits. This can
easily be understood as an artefact of an inconsistent target velocity: when gas parcels obtain some negative radial velocity for example as
they are influenced by planet-driven spiral waves, the damping of only the azimuthal velocity component leads to modifications of potentially
eccentric orbits.
While the initial migration of the Neptune looks similar to the reference case,
the resonance of the giant pair is affected by this modification to the inner disk and the gap. In this model, they do not settle into
a deeply resonant state, but only one resonant angle librates which is accompanied by a high
apsidal precession rate of the inner planet $\dot{\varpi}_\mathrm{c}$. This altered behaviour can be explained by a significantly lower
eccentricity of the innermost planet c, that remains low at around $e_\mathrm{c} \sim 0.05$ instead of increasing to $\sim 0.2$,
as is the case in our reference model. As argued by \citet{Kley2005}, this results in a higher resonance-induced precession rate
of the inner planet due to the inverse dependence $\dot{\varpi}_j \sim 1/e_j$ for small $e_j$ in a two-body resonance.
In this configuration, the Neptune is able to remain trapped in the 3:1 MMR, with one resonant angle librating most of the time.
This shows how an unrealistic treatment of the inner boundary can even affect the evolution of the outermost planet in an indirect way by
a more efficient damping of the eccentricity of the innermost planet.
Compared to the \texttt{a3-h.05-.25s} model, torque variations are similarly periodic, but their amplitude is smaller. This allows for trapping
in resonance, which did not occur in the other case.
Regarding the disk eccentricity, we observe how, as the Neptune remains further away from the gap
edge, this part of the disk is able to maintain a substantial eccentricity, as opposed to the reference model, where the planet migrates inward
and circularizes it. The inner disk that is significantly depleted in mass, shows higher eccentricities.
We discard this damping prescription as unphysical and inconsistent and caution that, although initially its effect appears to be restricted
to the cavity in the inner disk, it has significant implications on the evolution of the system.

Significantly improved agreement with an undamped model, that suffers from a smaller inner cavity,
can be achieved by additionally damping the radial velocity towards viscous speeds, which we chose for our reference setups.
In a model that uses this viscous outflow prescription at the inner boundary with $\beta = 100/3$, which is ten times higher than
the reference value, we find only a slight decrease of the order one in the surface density of the inner disk and dynamically behaves
almost identical, suggesting that as long as gas is not lost in an artificially created cavity, the balance of accretion onto the
inner disk and gas removal close to the inner boundary leads to a similar structure and damping properties of the inner disk.

The pink dotted lines in Fig. \ref{fig:sigrel_app} show results for a model that uses the same damping towards viscous velocities, but
allows for no flux through the inner boundary, which is reflective. The initial structure of the model closely follows that of the reference
model, showing a slight increase of the surface density in the inner disk.
As the simulations evolve with time, the inner disk and the
joint gap remain at higher densities, while they are depleted in the reference model.
The higher density in the gap results in an increased migration rate of the giant planets.
While, the Neptune is able to approach the giant planets, it stops getting closer just outside the 2:1 MMR. This is possibly related to the fact
that the giant planets are migrating at a higher rate in this model.
We also note that the eccentricity of the innermost planet remains higher, as it is still being driven inward.

\section{Closer view of the inner disk}
\label{sec:AppB}
\begin{figure}
   \centering
   \includegraphics[width=.5\textwidth]{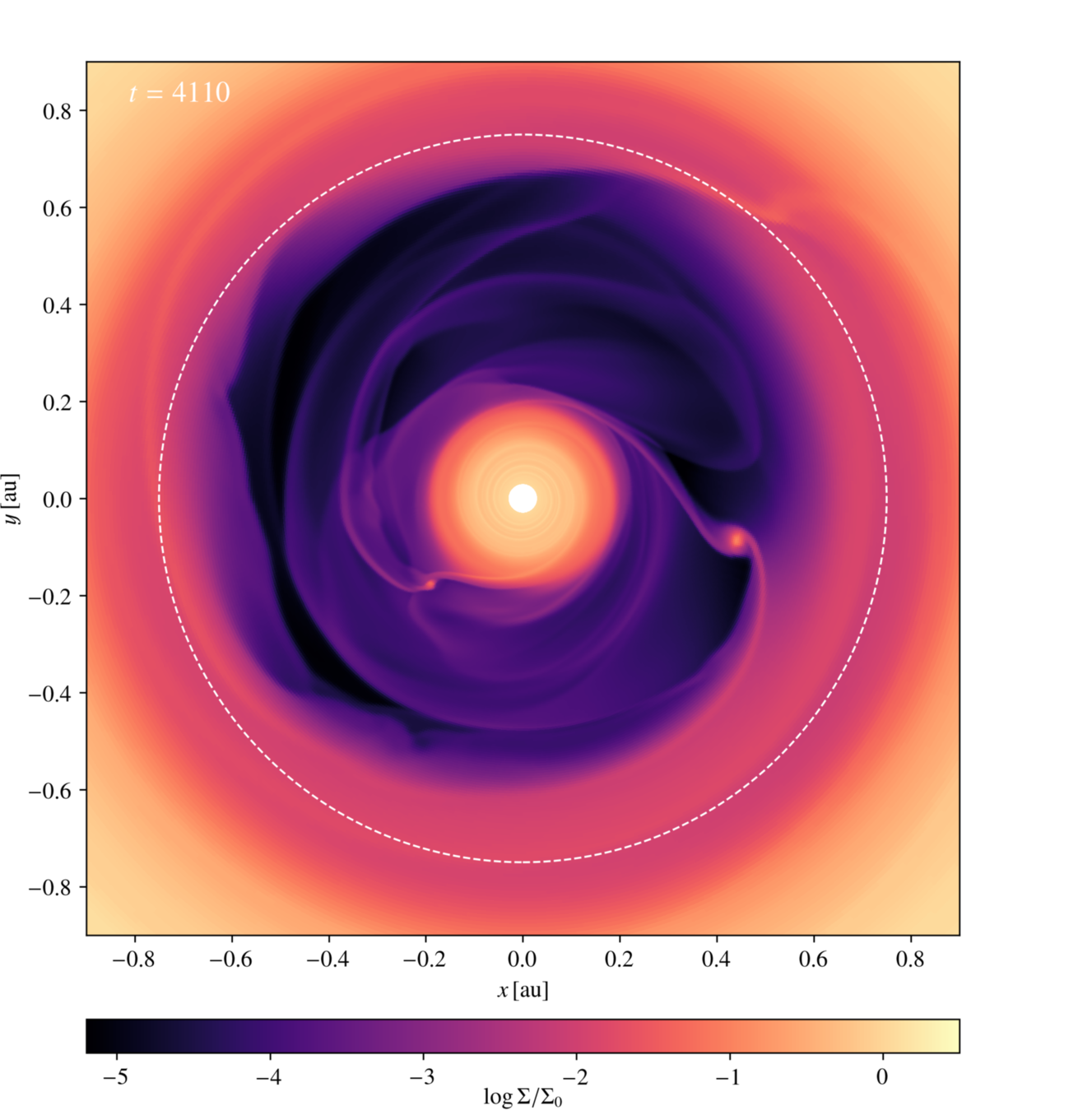}
		\caption{Magnified version of Fig \ref{fig:sig2dRef}. We note that the range of the colourbar has been slightly adjusted to improve the
		visibility.}
           \label{fig:sig2dRefZoom}
\end{figure}

In Fig. \ref{fig:sig2dRefZoom} we show a zoomed-in version of the two-dimensional surface-density distribution for
the reference simulation (Fig. \ref{fig:sig2dRef}).
The structure and evolution of the inner disk and the gap are easier to recognize in this view.
It is clearly visible that the damping prescription only affects the very inner part of the inner disk.

\section{A model in higher resolution}
\label{sec:AppRes}
\begin{figure}
           \includegraphics[width=.5\textwidth]{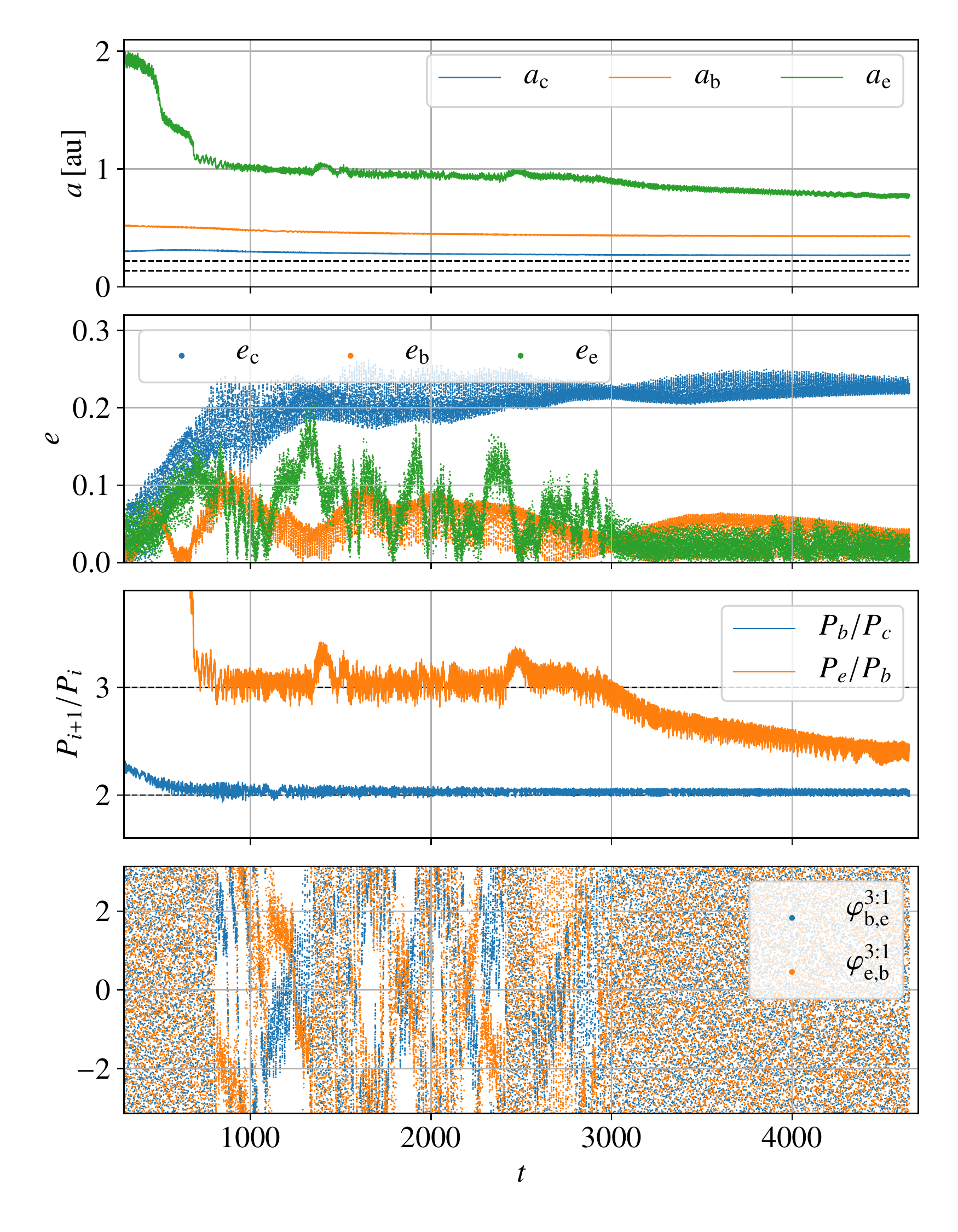}
   \vspace*{-2.5em}
   \caption{Time evolution of the orbital elements for a higher resolution model. We note that in this model the Neptune does not reverse its
   migration and is temporarily engaged in a 3:1 MMR with planet b.
   }
   \label{fig:HRMig}
\end{figure}
We conducted a resolution study by performing a simulation with our reference parameters with twice the number of cells in both
dimensions, such that $N_r \times N_\varphi = 1152 \times 1416$. The resulting orbital evolution of the planets is displayed in
Fig. \ref{fig:HRMig}. While the inner planet pair behaves virtually identical, we notice that the outer planet does not reverse its
migration after getting close to the 3:1 MMR, but is trapped into the resonance.
What exactly causes this discrepancy is still unclear and requires further investigation.
When it obtains substantial eccentricity at around $t \simeq 1300$ and $t \simeq 2500$, the resonance is briefly broken and then
re-established. Around $t \simeq 3000$, the Neptune passes the 3:1 MMR and migrates into the gap, as is the case for the reference model.
Due to the high computational cost, we were not able to evolve this simulation further. Thus, it is still unclear if the slight differences we
see here will become important in the future evolution of this simulation.

\end{appendix}

\end{document}